\documentclass{article}
\usepackage{amsmath, amsthm, amssymb, graphicx, hyperref, listings, textcomp}
\usepackage[margin=1in]{geometry}
\DeclareMathOperator{\diag}{diag}
\begin{document}
\setlength{\unitlength}{1cm}
\title{An Introduction to Chameleon Gravity}
\author{T. P. Waterhouse}
\date{22 November 2006}
\maketitle
\begin{abstract}
Following work by Khoury and Weltman, we introduce a scalar field $\phi$, the chameleon, which is conformally coupled to matter.  That is, matter experiences a metric which is a conformal transform (parametrized by $\phi$) of the Einstein metric.  The effective potential of the field $\phi$ is a sum of its self-interaction term and an exponential term due to the conformal coupling.  Under certain conditions on the self-interaction and the coupling, this effective potential has a minimum which depends on the local matter density, as does its second derivative at the minimum.  As a result, the scalar field acquires a mass which increases with local matter density.

The field $\phi$ mediates a fifth force which is suppressed in the laboratory and in interactions between large bodies like planets, but which may be detectable between small test masses in space.

In this pedagogical essay, we derive the equation of motion of $\phi$, discuss chameleon-field cosmology, and examine some simple solutions with a view to experimental detection of the chameleon.
\end{abstract}
\tableofcontents
\lstlistoflistings
\listoffigures
\section{Introduction}
\subsection{Why chameleons?}
One of the biggest surprises in modern cosmology was the discovery that the expansion of the Universe is accelerating.  Taken together with other theoretical and observational evidence \cite{Carroll}, this has been interpreted to imply that Einstein's cosmological constant $\Lambda$, or some other mechanism for acceleration, must feature prominently in cosmology.

With the Universe spatially flat or very nearly so, the energy density $\rho_\Lambda$ associated with a cosmological constant (or more generally, with dark energy) is of order $\rho_\text{critical}\equiv3H_0^2/8\pi G$ today.  But in the cosmological constant scenario, $\rho_\Lambda$ is constant in time, which means that just after inflation, $\rho_\Lambda$ and $\rho_r$, the energy density of radiation, differed by over 100 orders of magnitude \cite{0210533}.  This fine-tuning, and other conceptual problems with the cosmological constant, have been discussed extensively in the literature, for example in \cite{Carroll, Weinberg}.

A more general model for cosmic acceleration is a slowly-rolling scalar field, called quintessence (see \cite{0403324} for a recent review).  The slow roll means that the field has negative pressure and therefore acts to accelerate expansion.  Quintessence can resolve the fine-tuning problem via tracker potentials, couplings between the quintessence field and dark matter, or modifications to general relativity.

Scalar fields, for example the dilaton field, arise necessarily in string theory and are generally coupled to matter.  But if such a field couples strongly to matter, we should have detected it by now as a ``fifth force''.  A mechanism must therefore exist to suppress the detectable effect of any coupling to matter.

Khoury and Weltman \cite{KW} have suggested a coupling which gives the scalar field a mass depending on the local density of matter.  In regions of high density (like the Earth), the field has a large mass; in regions of low density (like interstellar space), the field has a small mass.  In terrestrial experiments, the large mass of the field suppresses its interaction with matter.  In observations of the Solar System, the action of the field is suppressed by a ``thin-shell mechanism'', to be discussed.  Since the field is able to hide so well from our observations and experiments, it is called a chameleon.

A specific example of a chameleon field arising from string theory is given in \cite{BvdBD}, which shows that under certain conditions, the radion field of brane theory can act as a chameleon.

Chameleon fields are not limited to quintessence.  Chameleon behaviour can be exhibited in, for example, $\phi^4$ theory, using the mechanism we describe below but with the sign of the coupling constants $\beta_i$ reversed \cite{GK, UGK}.
\subsection{Organization of this essay}
This article is intended to be a pedagogical introduction to chameleon gravity.  For a complementary and more formal review, see Mota and Shaw \cite{MS}.

We will begin by introducing the action for gravity, matter, and the chameleon field $\phi$, which we will use to determine the equation of motion for $\phi$.  We will find that we can write down an equation of motion for $\phi$ in terms of a bare quintessence-type potential $V$ and a contribution from the matter fields, which will give rise to a local relationship between the mass of the chameleon field and the density of matter.

With this in mind, we will then briefly detour into a discussion of the potential $V$, finding that an exponential potential allows chameleon behaviour while causing cosmic acceleration, but that a power-law potential cannot cause acceleration without violating, by several orders of magnitude, existing experimental bounds for laboratory detection.

Next, we will discuss the cosmological evolution of the chameleon field.  The principal result of this section will be that $\phi$ keeps up with an attractor solution $\phi_\text{min}$, analogous to tracker solutions in quintessence, from the very early universe until today.

Then we will derive the equation for the chameleon force acting on a test mass in a static chameleon field.  A lengthy derivation will give us an approximate solution for static spherically symmetric chameleon fields which can be condensed into a useful Maple script.  The important result of this section will be a ``thin-shell suppression factor'' which explains why large objects like the Earth don't give rise to large chameleon forces.

Finally, we will use our solution to show explicitly both why the chameleon field may have eluded detection so far and why it may potentially be observed by satellite experiments.
\subsection{Conventions}
In this essay we will set $c=1$ and $\hbar=1$ and use the metric signature $(-,+,+,+)$.  We will denote by $M_\text{pl}$ the reduced Planck mass:
\begin{displaymath}
M_\text{pl}\equiv\left(8\pi G\right)^{-1/2}\approx2.44\times10^{18}\,\text{GeV}.
\end{displaymath}

Since we are dealing with what is essentially a scalar-tensor theory of gravitation, we will be making use of the terms ``Jordan frame'' and ``Einstein frame''.  The Einstein frame uses the bare spacetime metric $g_{\mu\nu}$, whereas the Jordan frame uses a metric $g_{\mu\nu}^{(i)}$ which is conformally related to $g_{\mu\nu}$ (the index $i$ may be used to enumerate matter species, allowing a different coupling for different species).  The importance of the Jordan frame is that in a scalar-tensor theory, $g_{\mu\nu}^{(i)}$ is the metric which matter experiences.  In what follows, we will make clear which frame goes with which equation.  See the first three chapters of \cite{FM} for an introduction to scalar-tensor theories and conformal transformations.

For simplicity of notation, we will often write $\tilde{X}$ in place of $X_{(i)}$ when we are working with a fixed index $i$.
\section{The behaviour of the chameleon field}
\subsection{The action and its variation for a chameleon field}
Recall the normalised vacuum Einstein-Hilbert action
\begin{displaymath}
S_{\text{EH}}=\int d^4 x\sqrt{-g}\frac{1}{16\pi G}R=\int d^4 x\sqrt{-g}\frac{M_\text{pl}^2}{2}R.
\end{displaymath}
Introduce a scalar field $\phi$ with potential $V(\phi)$ and action
\begin{displaymath}
S_\phi=-\int d^4 x\sqrt{-g}\left\{\frac{1}{2}(\partial\phi)^2+V(\phi)\right\}.
\end{displaymath}
So far, this is the standard picture.  But now we introduce a set of matter fields $\psi_\text{m}^{(i)}$ with action
\begin{displaymath}
S_\text{m}=-\int d^4 x \mathcal{L}_\text{m}\left(\psi_\text{m}^{(i)},g_{\mu\nu}^{(i)}\right),
\end{displaymath}
which are coupled to $\phi$ by the definition
\begin{equation}
g_{\mu\nu}^{(i)}\equiv e^{2\beta_i\phi/M_\text{pl}}g_{\mu\nu},
\label{conformalrelation}
\end{equation}
where $\beta_i$ are dimensionless coupling constants, in principle one for each matter species.  Note that $\mathcal{L}$ is a scalar density of weight 1 in the language of differential geometry.

The total action including gravity, $\phi$, and matter is thus\footnote{Although we introduced $\phi$ as a rolling scalar field which happens to interact with matter, we could alternately have come up with $S$ as the action of a Brans-Dicke theory in which the scalar field happens to have a self-interaction potential $V\left(\phi\right)$.  In the case that $i$ takes only one value, we could also have written $S$ in terms of the metric $g_{\mu\nu}^{(i)}$, making the role of $\phi$ as a fifth force clearer.  See \cite[pp.38--42;66--8]{FM}, for more on both these points.}
\begin{equation}
\label{total_action}
\begin{aligned}
S&=\int d^4 x\sqrt{-g}\left\{\frac{M_\text{pl}^2}{2}R-\frac{1}{2}\left(\partial\phi\right)^2-V(\phi)\right\}-\int d^4 x \mathcal{L}_\text{m}\left(\psi_\text{m}^{(i)},g_{\mu\nu}^{(i)}\right)\\
&=\int d^4 x\sqrt{-g}\left\{\frac{M_\text{pl}^2}{2}R-\frac{1}{2}\nabla_\mu\phi\nabla^\mu\phi-V(\phi)-\frac{1}{\sqrt{-g}}\mathcal{L}_\text{m}\left(\psi_\text{m}^{(i)},g_{\mu\nu}^{(i)}\right)\right\}.
\end{aligned}
\end{equation}
Variation with respect to $\phi$ allows us to obtain the equation of motion for $\phi$:
\begin{align*}
\delta S&=\int d^4 x\sqrt{-g}\left\{-\nabla_\mu\phi\delta\nabla^\mu\phi-V_{,\phi}(\phi)\delta\phi-\frac{1}{\sqrt{-g}}\frac{d\mathcal{L}_\text{m}}{d\phi}\delta\phi\right\}\\
&=\int d^4 x\sqrt{-g}\left\{-\nabla_\mu\phi\nabla^\mu\delta\phi-V_{,\phi}(\phi)\delta\phi-\sum_i\frac{1}{\sqrt{-g}}\frac{\partial\mathcal{L}_\text{m}}{\partial g_{\mu\nu}^{(i)}}\frac{\partial g_{\mu\nu}^{(i)}}{\partial\phi}\delta\phi\right\}\\
&=\int d^4 x\sqrt{-g}\left\{\left(\nabla_\mu\nabla^\mu\phi\right)\delta\phi-V_{,\phi}(\phi)\delta\phi-\sum_i\frac{1}{\sqrt{-g}}\frac{\partial\mathcal{L}_\text{m}}{\partial g_{\mu\nu}^{(i)}}\frac{2\beta_i}{M_\text{pl}}g_{\mu\nu}^{(i)}\delta\phi\right\}\\
&=\int d^4 x\sqrt{-g}\left\{\nabla^2\phi-V_{,\phi}(\phi)-\sum_i\frac{1}{\sqrt{-g}}\frac{\partial\mathcal{L}_\text{m}}{\partial g_{\mu\nu}^{(i)}}\frac{2\beta_i}{M_\text{pl}}g_{\mu\nu}^{(i)}\right\}\delta\phi
\end{align*}
The manipulations of the kinetic term in this derivation deserve careful explanation.  On the first line we used the symmetry of the metric to get $\delta\left(-\frac{1}{2}\nabla_\mu\phi\nabla^\mu\phi\right)=-\nabla_\mu\phi\delta\nabla^\mu\phi$.  On the second line we used the commutativity of differentiation with variation to get $\delta\nabla^\mu\phi=\nabla^\mu\delta\phi$.  On the third line we integrated by parts by applying the divergence theorem to $\sqrt{-g}\left(\nabla^\mu\phi\right)\delta\phi$, which is a vector density of weight 1, and assuming that $\nabla^\mu\phi\rightarrow 0$ or $\delta\phi\rightarrow 0$ at spacetime infinity or boundaries.

Requiring $\delta S=0$ we thus have an equation of motion:
\begin{equation}
\nabla^2\phi=V_{,\phi}(\phi)+\sum_i\frac{1}{\sqrt{-g}}\frac{\partial\mathcal{L}_\text{m}}{\partial g_{\mu\nu}^{(i)}}\frac{2\beta_i}{M_\text{pl}}g_{\mu\nu}^{(i)}.
\label{firsteom}
\end{equation}
But the sum on the right depends on both the matter content and $\phi$.  We would like to identify and isolate the matter content so that we can write an equation of motion with manifest $\phi$-dependence.
\subsection{Defining the energy density of matter}
\label{Defining the energy density of matter}
In this subsection, we will fix $i$ and write $\tilde{X}$ in place of $X_{(i)}$.
\subsubsection{Jordan frame}
Assuming that the matter fields $\psi_\text{m}^{(i)}$ do not interact with each other \cite{KW}, each energy-momentum tensor
\begin{displaymath}
\tilde{T}^{\mu\nu}\equiv-\frac{2}{\sqrt{-\tilde{g}}}\frac{\partial\mathcal{L}_\text{m}}{\partial\tilde{g}_{\mu\nu}}
\end{displaymath}
is conserved in the Jordan frame \cite[pp.38--42]{FM}.  That is to say,
\begin{equation}
\widetilde{\nabla}_{\nu}\tilde{T}^{\mu\nu}=0,
\label{conservationequation}
\end{equation}
where $\widetilde{\nabla}$ is the metric connection corresponding to the metric $\tilde{g}_{\mu\nu}$.

Assume the matter to be a perfect isentropic fluid with $\tilde{p}=w_i\tilde{\rho}$.  Then
\begin{equation}
\tilde{T}^{\mu\nu}\tilde{g}_{\mu\nu}=-\tilde{\rho}+3\tilde{p}=-\left(1-3w_i\right)\tilde{\rho}.
\label{densityandpressure}
\end{equation}
This is just the standard form, as in \cite[p.140]{MTW}, written in the Jordan frame.

Note that since matter of species $i$ couples universally to the metric $\tilde{g}_{\mu\nu}$, experiments composed of matter of species $i$ make measurements in terms of $\tilde{g}_{\mu\nu}$ \cite{Damour}; in particular, the energy density which we measure is $\tilde{\rho}$.
\subsubsection{Einstein frame}
Impose, without loss of generality, a Friedmann-Lema\^itre-Robertson-Walker (FLRW) background metric.  The energy density $\rho$ in the Einstein frame (the frame corresponding to $g_{\mu\nu}$) is that quantity, conformally related to $\tilde{\rho}$, which obeys the standard continuity equation $\rho\propto a^{-3\left(1+w_i\right)}$.

We will need the following:
\begin{gather*}
\tilde{a}\equiv e^{\beta_i\phi/M_\text{pl}}a;\\
\begin{aligned}
\tilde{g}_{\mu\nu}=e^{2\beta_i\phi/M_\text{pl}}g_{\mu\nu}&=e^{2\beta_i\phi/M_\text{pl}}\diag\left(-1,a^2,a^2,a^2\right)\\
&=\diag\left(-e^{2\beta_i\phi/M_\text{pl}},\tilde{a}^2,\tilde{a}^2,\tilde{a}^2\right);
\end{aligned}
\\
\tilde{g}^{\mu\nu}=\diag\left(-e^{-2\beta_i\phi/M_\text{pl}},\tilde{a}^{-2},\tilde{a}^{-2},\tilde{a}^{-2}\right).
\end{gather*}

Compute the Christoffel symbols:
\begin{align*}
\tilde{\Gamma}_{03}^{3}=\tilde{\Gamma}_{02}^{2}=\tilde{\Gamma}_{01}^{1}&=\frac{1}{2}\tilde{g}^{1\alpha}\left(\tilde{g}_{\alpha0,1}+\tilde{g}_{\alpha1,0}-\tilde{g}_{01,\alpha}\right)=\frac{1}{2}\tilde{g}^{11}\left(\tilde{g}_{10,1}+\tilde{g}_{11,0}-\tilde{g}_{01,1}\right)\\
&=\frac{1}{2}\tilde{a}^{-2}\left(0+2\tilde{a}\tilde{a}_{,0}-0\right)=\tilde{a}^{-1}\tilde{a}_{,0};\\
\tilde{\Gamma}_{33}^{0}=\tilde{\Gamma}_{22}^{0}=\tilde{\Gamma}_{11}^{0}&=\frac{1}{2}\tilde{g}^{0\alpha}\left(2\tilde{g}_{\alpha1,1}-\tilde{g}_{11,\alpha}\right)=\frac{1}{2}\tilde{g}^{00}\left(2\tilde{g}_{01,1}-\tilde{g}_{11,0}\right)\\
&=-\frac{1}{2}e^{-2\beta_i\phi/M_\text{pl}}\left(0-2\tilde{a}\tilde{a}_{,0}\right)=e^{-2\beta_i\phi/M_\text{pl}}\tilde{a}\tilde{a}_{,0}.
\end{align*}
Then use them to expand the conservation equation (\ref{conservationequation}) with index $\mu=0$, remembering when we differentiate that we are holding $\phi$ fixed and simply varying the scale factor $a$:
\begin{align*}
0&=\widetilde{\nabla}_{\nu}\tilde{T}^{0\nu}\\
&=\tilde{T}_{,\nu}^{0\nu}+\tilde{\Gamma}_{\sigma\nu}^{0}\tilde{T}^{\sigma\nu}+\tilde{\Gamma}_{\sigma\nu}^{\nu}\tilde{T}^{0\sigma}\\
&=\tilde{T}_{,0}^{00}+\tilde{\Gamma}_{\sigma\nu}^{0}\tilde{T}^{\sigma\nu}+\tilde{\Gamma}_{0\nu}^{\nu}\tilde{T}^{00}\\
&=\left(-\tilde{\rho}\tilde{g}^{00}\right)_{,0}+3e^{-2\beta_i\phi/M_\text{pl}}\tilde{a}\tilde{a}_{,0}\tilde{p}\tilde{g}^{11}+3\tilde{a}^{-1}\tilde{a}_{,0}\left(-\tilde{\rho}\tilde{g}^{00}\right)\\
&=e^{-2\beta_i\phi/M_\text{pl}}\left(\tilde{\rho}_{,0}+3\tilde{a}^{-1}\tilde{a}_{,0}\tilde{p}+3\tilde{a}^{-1}\tilde{a}_{,0}\tilde{\rho}\right)\\
&=e^{-2\beta_i\phi/M_\text{pl}}\left(\tilde{\rho}_{,0}+3\left(1+w_i\right)\tilde{a}^{-1}\tilde{a}_{,0}\tilde{\rho}\right).
\end{align*}
Thus, multiplying by $e^{2\beta_i\phi/M_\text{pl}}\tilde{a}^{3\left(1+w_i\right)}$:
\begin{align*}
0&=\tilde{a}^{3\left(1+w_i\right)}\tilde{\rho}_{,0}+3\left(1+w_i\right)\tilde{a}^{3\left(1+w_i\right)-1}\tilde{a}_{,0}\tilde{\rho}\\
&=\left(\tilde{a}^{3\left(1+w_i\right)}\tilde{\rho}\right)_{,0}\\
&=\left(a^{3\left(1+w_i\right)}e^{3\left(1+w_i\right)\beta_i\phi/M_\text{pl}}\tilde{\rho}\right)_{,0}.
\end{align*}
In other words, in the Einstein frame, the quantity
\begin{equation}
\label{Einstein_density}
\rho\equiv e^{3\left(1+w_i\right)\beta_i\phi/M_\text{pl}}\tilde{\rho}
\end{equation}
is the energy density of matter, because it obeys the continuity equation $\rho\propto a^{-3\left(1+w_i\right)}$.  (Of course, if $\phi\ll M_\text{pl}$, then $\rho\approx\tilde\rho$.)

Restoring the $i$ subscripts and using equation (\ref{densityandpressure}), we have the Einstein-frame energy density for each species $i$:
\begin{equation}
\begin{aligned}
\rho_i&=-e^{3\left(1+w_i\right)\beta_i\phi/M_\text{pl}}\frac{1}{1-3w_i}T_{(i)}^{\mu\nu}g_{\mu\nu}^{(i)}\\
&=e^{3\left(1+w_i\right)\beta_i\phi/M_\text{pl}}\frac{1}{1-3w_i}\frac{2}{\sqrt{-g^{(i)}}}\frac{\partial\mathcal{L}_\text{m}}{\partial g_{\mu\nu}^{(i)}}g_{\mu\nu}^{(i)}\\
&=e^{3\left(1+w_i\right)\beta_i\phi/M_\text{pl}}\frac{1}{1-3w_i}\frac{2}{\sqrt{e^{8\beta_i\phi/M_\text{pl}}\left(-g\right)}}\frac{\partial\mathcal{L}_\text{m}}{\partial g_{\mu\nu}^{(i)}}g_{\mu\nu}^{(i)}\\
&=e^{-\left(1-3w_i\right)\beta_i\phi/M_\text{pl}}\frac{1}{1-3w_i}\frac{2}{\sqrt{-g}}\frac{\partial\mathcal{L}_\text{m}}{\partial g_{\mu\nu}^{(i)}}g_{\mu\nu}^{(i)}.
\end{aligned}
\label{Einstein_energydensity}
\end{equation}
\subsection{The chameleon equation of motion}
\label{sectioncontainingbetaunity}
Substituting equation (\ref{Einstein_energydensity}) into equation (\ref{firsteom}), we obtain an equation of motion in which all the $\phi$ dependence is explicit:
\begin{displaymath}
\nabla^2\phi=V_{,\phi}(\phi)+\sum_i\left(1-3w_i\right)\frac{\beta_i}{M_\text{pl}}\rho_i e^{\left(1-3w_i\right)\beta_i\phi/M_\text{pl}}.
\end{displaymath}

We may thus express the dynamics for $\phi$ in terms of a single effective potential,
\begin{equation}
\label{veff}
V_{\text{eff}}(\phi)\equiv V(\phi)+\sum_i\rho_i e^{\left(1-3w_i\right)\beta_i\phi/M_\text{pl}}.
\end{equation}
Then the chameleon equation of motion in the Einstein frame is simply
\begin{equation}
\nabla^2\phi=V_{\text{eff},\phi}(\phi).
\label{conciseeom}
\end{equation}
Recalling our choice of metric signature, this means that we expect $\phi$ to seek out minima of $V_{\text{eff}}$.

Note that in the case of non-relativistic matter, $w_i\approx0$, so the effective potential simplifies to
\begin{displaymath}
V_{\text{eff}}(\phi)=V(\phi)+\sum_i\rho_i e^{\beta_i\phi/M_\text{pl}}.
\end{displaymath}
We will later be assuming $\beta_i=1$, since we wish these coupling constants to take on ``natural'' values of order unity.  If all $\beta_i$ are equal, then we may also assume we are working with a single matter species.
\subsection{The potential and the effective potential}
\label{potentialandeffectivepotential}
We wish to choose a bare potential $V\left(\phi\right)$ which can give rise to cosmic acceleration today via the slow-roll mechanism, as in quintessence \cite{Ratra_Peebles}.  It would be cumbersome to come up with a mechanism that makes $\phi$ act as a cosmological constant only today, so we will assume that it has always been rolling down a potential slope in, without loss of generality, the positive direction.  Thus $V\left(\phi\right)$ should be a monotonically decreasing function of $\phi$.

To obtain chameleon behaviour, we will see below that it is sufficient to assume that $V\left(\phi\right)$ is of the ``runaway'' form, in the following sense:
\begin{enumerate}
\item$\lim_{\phi\rightarrow0}V(\phi)=\infty$;
\item$V(\phi)$ is $C^\infty$, bounded below, and strictly deceasing;
\item$V_{,\phi}(\phi)$ is strictly negative and increasing;
\item$V_{,\phi\phi}(\phi)$ is strictly positive and decreasing.
\end{enumerate}
We concomitantly confine $\phi$ to positive values.  Note that although runaway potentials are not found in classical physics, they do arise in supersymmetric models; see \cite{PB} for a discussion of such potentials in the context of quintessence.

A convenient example which is often seen in quintessence models (see \cite{ZWS}, for example) is the inverse power-law potential (also called the Ratra-Peebles potential),
\begin{displaymath}
V(\phi)=\frac{M^{4+n}}{\phi^n},
\end{displaymath}
where $M$ is a constant with the dimension of mass and $n$ is a positive constant.

Another example is the exponential potential
\begin{displaymath}
V(\phi)=M^4\exp\left(\frac{M^n}{\phi^n}\right),
\end{displaymath}
where again $M$ is a constant mass and $n$ is positive.

The important difference between these potential functions is in the limit $\lim_{\phi\rightarrow\infty}V(\phi)$.  The Ratra-Peebles potential goes to 0, whereas the exponential potential goes to $M^4$.  Therefore, present-day observations of the acceleration of the expansion of the Universe give estimates of the value of the constant $M$ which differ by several orders of magnitude between the two potentials, as we will see later.  This is brought to the reader's attention now to avoid later confusion, since Khoury and Weltman \cite{KW} and Brax et al.\ \cite{BvdBDKW} discuss both models.

\begin{figure}[ht]
\centering
\includegraphics[scale=0.43,angle=270]{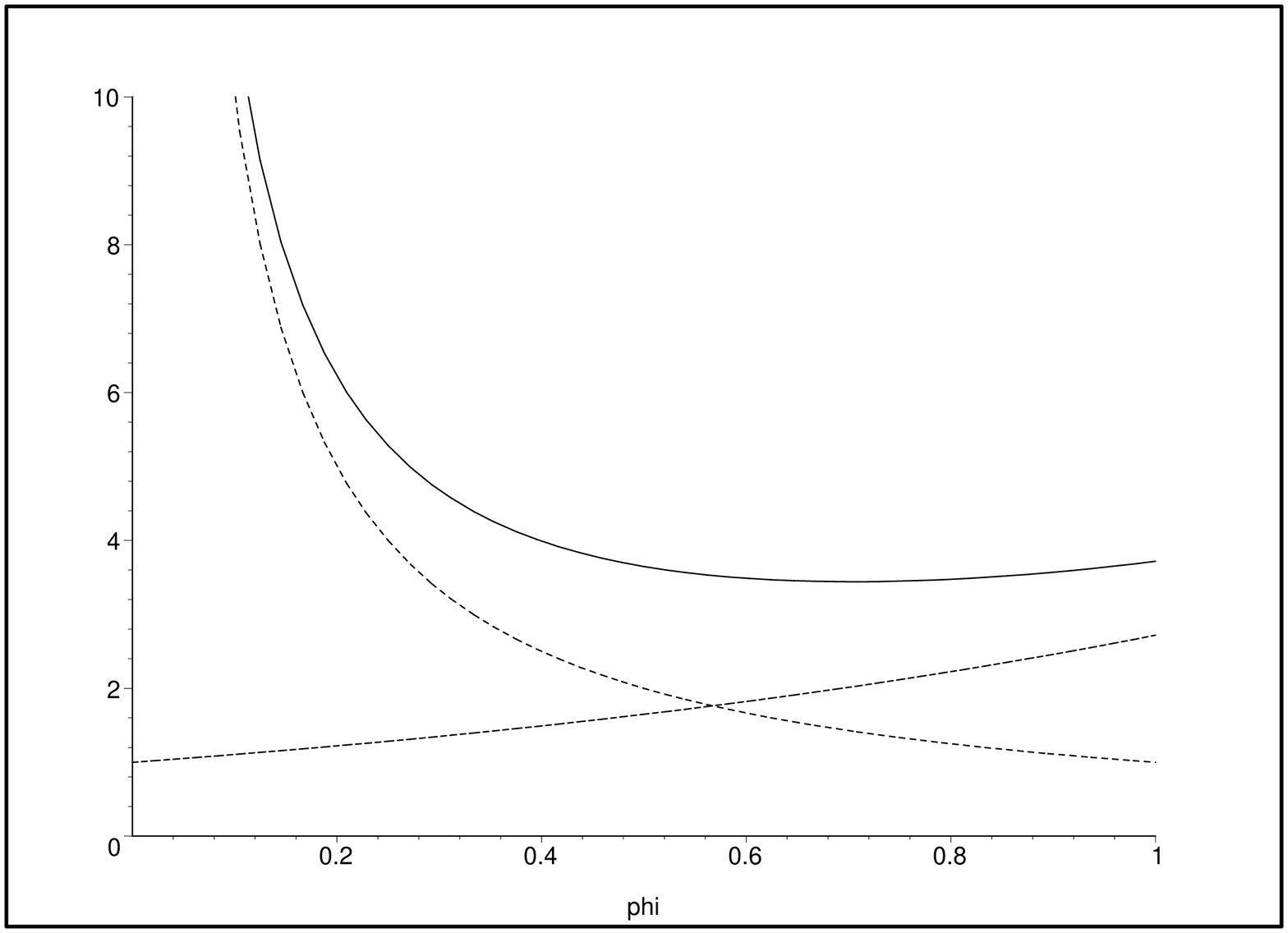}
\includegraphics[scale=0.43,angle=270]{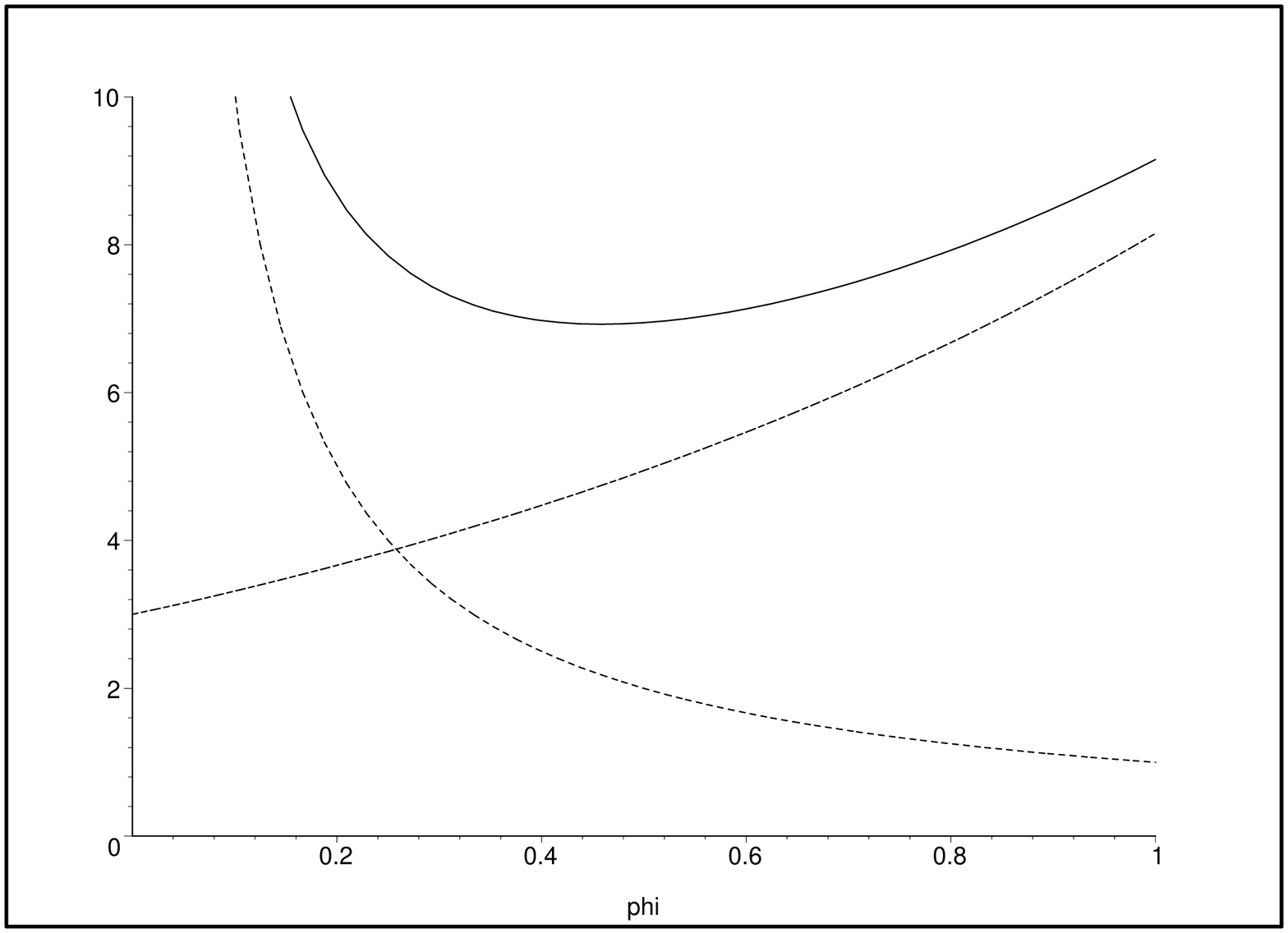}
\caption[$\rho e^{\phi}$, $V(\phi)$, and $V_{\text{eff}}(\phi)$]{A simple example showing the behaviour of physically relevant quantities, with units and constants suppressed.  Specifically, we plot $\rho e^{\phi}$ (dashed line), $V(\phi)=\phi^{-1}$ (dotted line), and their sum, $V_{\text{eff}}(\phi)$ (solid line).  $\rho=1$ in the first plot and $\rho=3$ in the second plot.}
\label{veffplot}
\end{figure}

If the coupling constants $\beta_i$ are positive, then although $V(\phi)$ is monotonically decreasing, the presence of matter gives the effective potential (\ref{veff}) a minimum at a finite field value $\phi=\phi_\text{min}$, which depends on $\rho_i$ (as in Figure \ref{veffplot}).  Note that $V_{\text{eff},\phi}\left(\phi_\text{min}\right)=0$, ie,
\begin{equation}
V_{,\phi}\left(\phi_\text{min}\right)+\sum_i\left(1-3w_i\right)\frac{\beta_i}{M_\text{pl}}\rho_i e^{\left(1-3w_i\right)\beta_i\phi_\text{min}/M_\text{pl}}=0.
\label{phimin}
\end{equation}
The mass associated with the field $\phi$ is given by
\begin{displaymath}
m^2\equiv V_{\text{eff},\phi\phi}\left(\phi\right)=V_{,\phi\phi}\left(\phi\right)+\sum_i\left(1-3w_i\right)^2\frac{\beta_i^2}{M_\text{pl}^2}\rho_i e^{\left(1-3w_i\right)\beta_i\phi/M_\text{pl}}.
\end{displaymath}
When $\phi=\phi_\text{min}$, we have $m^2=m_\text{min}^2$, where
\begin{equation}
m_\text{min}^2\equiv V_{,\phi\phi}\left(\phi_{\text{min}}\right)+\sum_i\left(1-3w_i\right)^2\frac{\beta_i^2}{M_\text{pl}^2}\rho_i e^{\left(1-3w_i\right)\beta_i\phi_\text{min}/M_\text{pl}}.
\label{mmin}
\end{equation}
The reason for this definition, as we will see later, is that $m_\text{min}$ is the inverse of the characteristic range of the chameleon force in a given medium.

Equation (\ref{phimin}) (see also Figure \ref{veffplot}) shows that if any density $\rho_i$ increases, then $\phi_\text{min}$ decreases, because $V_{,\phi}$ and $e^{\left(1-3w_i\right)\beta_i\phi/M_\text{pl}}$ are increasing functions of $\phi$ (it is reasonable to assume $w<\frac{1}{3}$).  Then we expect $m_\text{min}$ to increase, because $V_{,\phi\phi}$ is a decreasing function of $\phi$.  This is obvious from equation (\ref{mmin}) if $\phi_\text{min}\ll M_\text{pl}$, which we will later show to be the case for the chameleon model we will consider.

In the case where all the $\left(1-3w_i\right)\beta_i$ are equal (for example, if we are considering only one matter species), it is also easy to show that $m_\text{min}$ increases with each $\rho_i$, regardless of the magnitude of $\phi_\text{min}$.  If $\left(1-3w_i\right)\beta_i=B$ for each $i$, where $B$ is a positive constant, then equation (\ref{mmin}) becomes
\begin{align*}
m_\text{min}^2&=V_{,\phi\phi}\left(\phi_{\text{min}}\right)+\frac{B}{M_\text{pl}}\sum_i\left(1-3w_i\right)\frac{\beta_i}{M_\text{pl}}\rho_i e^{\left(1-3w_i\right)\beta_i\phi_\text{min}/M_\text{pl}}\\
&=V_{,\phi\phi}\left(\phi_{\text{min}}\right)-\frac{B}{M_\text{pl}}V_{,\phi}\left(\phi_\text{min}\right).
\end{align*}
Our conditions on $V\left(\phi\right)$ state that $V_{,\phi\phi}$ is a decreasing function of $\phi$ and $V_{,\phi}$ is an increasing function of $\phi$.  $\phi_\text{min}$ is a decreasing function of each $\rho_i$, and so $m_\text{min}$ is an increasing function of each $\rho_i$.

When does $\phi=\phi_\text{min}$?  We see from the equation of motion (\ref{conciseeom}) that if $V_{\text{eff},\phi}=0$, then $\nabla^2\phi=0$; for example, this is true for static, locally homogeneous solutions.  We will also see later that, for the exponential potential at least, $\phi\approx\phi_\text{min}$ cosmologically from very early times until today.
\subsection{The energy scale for the potential}
In this section, we will use the following WMAP 3-year data:  \cite{WMAP3}
\begin{gather*}
\Omega_\text{m}\approx0.237, \qquad \Omega_\text{DE}\approx0.763;\\
H_0\approx73.5\,\text{km}\,\text{s}^{-1}\,\text{Mpc}^{-1}.
\end{gather*}
We will also use
\begin{align*}
\rho_\text{critical}=\frac{3H_0^2}{8\pi G}&\approx4.37\times10^{-47}\,\text{GeV}^4;\\
\rho_\text{m}=\rho_\text{critical}\Omega_\text{m}&\approx1.04\times10^{-47}\,\text{GeV}^4;\\
\rho_\text{DE}=\rho_\text{critical}\Omega_\text{DE}&\approx3.34\times10^{-47}\,\text{GeV}^4.
\end{align*}
\subsubsection{The power-law potential}
\label{The power-law potential}
\begin{figure}[ht]
\centering
\includegraphics[scale=0.5,angle=270]{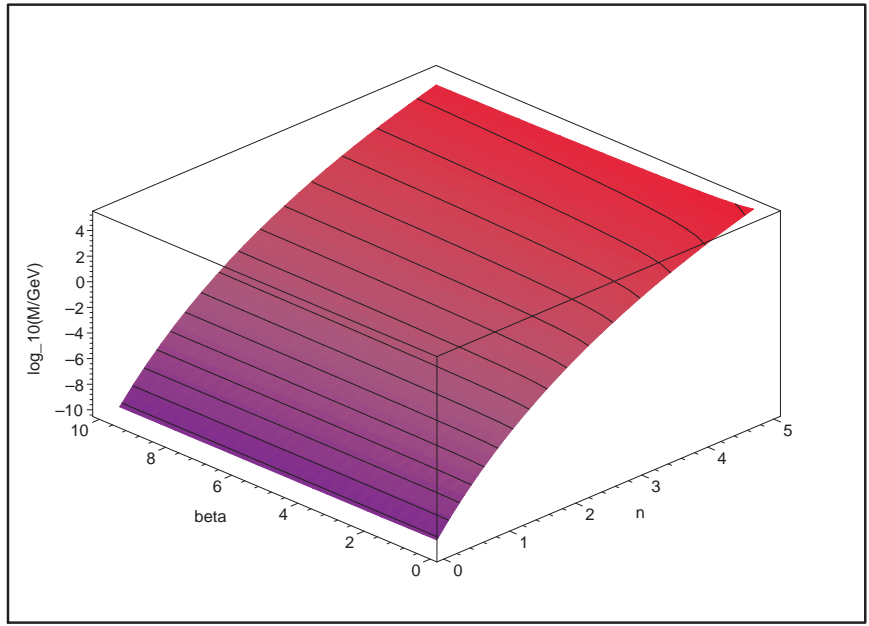}
\caption[Values of $M$ for the power-law potential]{The value of $\log_{10}M$, plotted as a function of $n$ and $\beta$, for the power-law potential.  Contours are drawn for integer values of $\log_{10}M$.}
\label{powerlaw_M}
\end{figure}
Suppose $V\left(\phi\right)=M^{4+n}\phi^{-n}$.  On cosmological scales, matter is described by a single species with $\tilde{\rho}=\rho_\text{m}$ and $w=0$.  We will denote by $\phi_\infty$ the value of $\phi_\text{min}$ corresponding to this density.  Assume that in the Universe at large, $\phi=\phi_\infty $ today; we will justify this assumption in section \ref{chameleoncosmology}.  Equation (\ref{phimin}), on substituting the relation (\ref{Einstein_density}), becomes
\begin{gather*}
-n\frac{M^{4+n}}{\phi_\infty^{n+1}}+\frac{\beta}{M_\text{pl}}e^{3\beta\phi_\infty/M_\text{pl}}\tilde{\rho}e^{\beta\phi_\infty/M_\text{pl}}=0\\
\Rightarrow \frac{\beta}{M_\text{pl}}\rho_\text{m}e^{4\beta\phi_\infty/M_\text{pl}}=n\frac{M^{4+n}}{\phi_\infty^{n+1}}.
\end{gather*}
For this field to account for the dark energy $\rho_\text{DE}$, we must now have $\rho_\text{DE}=V\left(\phi_\infty\right)=M^{4+n}\phi_\infty^{-n}$.  Substitution gives
\begin{gather*}
\frac{\beta}{M_\text{pl}}\rho_\text{m}e^{4\beta\phi_\infty/M_\text{pl}}=n\frac{\rho_\text{DE}\phi_\infty^n}{\phi_\infty^{n+1}}\\
\Rightarrow \frac{4\beta\phi_\infty}{M_\text{pl}}e^{4\beta\phi_\infty/M_\text{pl}}= 4n\frac{\Omega_\text{DE}}{\Omega_\text{m}}\\
\Rightarrow \phi_\infty=\frac{M_\text{pl}}{4\beta}W\left(4n\frac{1-\Omega_\text{m}}{\Omega_\text{m}}\right),
\end{gather*}
where $W$ is the Lambert (or product-log) function \cite{plog}.

For example, supposing $n=\beta=1$ and using $\Omega_\text{m}\approx0.237$, we get
\begin{displaymath}
\phi_\infty\approx 0.477\,M_\text{pl}\approx 1.16\times10^{18}\,\text{GeV}.
\end{displaymath}
Then $\rho_\text{DE}\approx3.34\times10^{-47}\,\text{GeV}^4$ gives
\begin{displaymath}
M=\left(\rho_\text{DE}\phi_\infty^n\right)^{1/\left(4+n\right)}\approx 2.08\,\text{keV}.
\end{displaymath}

In Figure \ref{powerlaw_M}, we explore the influence of $n$ and $\beta$ on the value of $M$ for the power-law potential.
\subsubsection{The exponential potential}
\label{The exponential potential}
Now suppose $V(\phi)=M^4\exp\left(M^n/\phi^n\right)$.  This potential is easy to analyse if we assume $\phi\gg M$ today, for then $\rho_\text{DE}\approx V(\phi)\approx M^4$, and so
\begin{displaymath}
M\approx\left(\rho_\text{DE}\right)^{1/4}\approx2.40\times10^{-3}\,\text{eV}.
\end{displaymath}

There is an important difference between the exponential potential and the power-law potential.  For large $\phi$ we have
\begin{displaymath}
M^4\exp\left(\frac{M^n}{\phi^n}\right)\approx M^4+\frac{M^{4+n}}{\phi^n}.
\end{displaymath}
That is, for $\phi\ll M$, the exponential potential and the power-law potential differ by the constant $M^4$.  Because this constant comes to dominate in the case of the exponential potential, this model suffers the same fine-tuning problem as does a bare cosmological constant $\Lambda$.  By contrast, the power-law potential is ``tuned'' by the interaction with matter, suggesting it to be a more natural choice.
\subsection{The chameleon in hiding}
\label{chameleoninhiding}
\begin{figure}[ht]
\centering
\includegraphics[scale=0.5,angle=270]{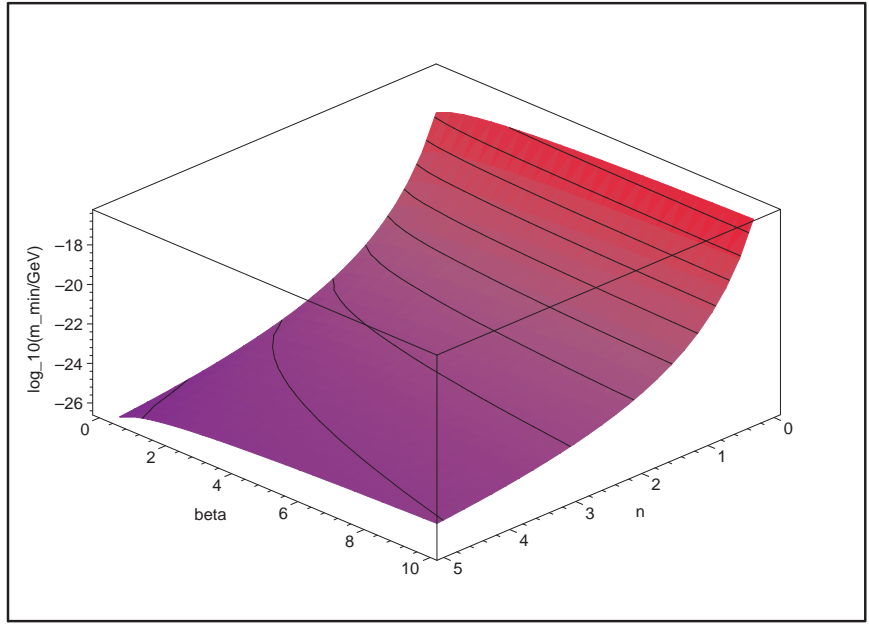}
\caption[Values of $m_\text{min}$ for the power-law potential]{The value of $\log_{10}m_\text{min}$ in Earth's atmosphere, plotted as a function of $n$ and $\beta$, for the power-law potential.  Contours are drawn for integer values of $\log_{10}m_\text{min}$.  Note that the $n$ and $\beta$ axes are inverted here, relative to Figures \ref{powerlaw_M} and \ref{exponential_mmin}.}
\label{powerlaw_mmin}
\end{figure}
\begin{figure}[ht]
\centering
\includegraphics[scale=0.5,angle=270]{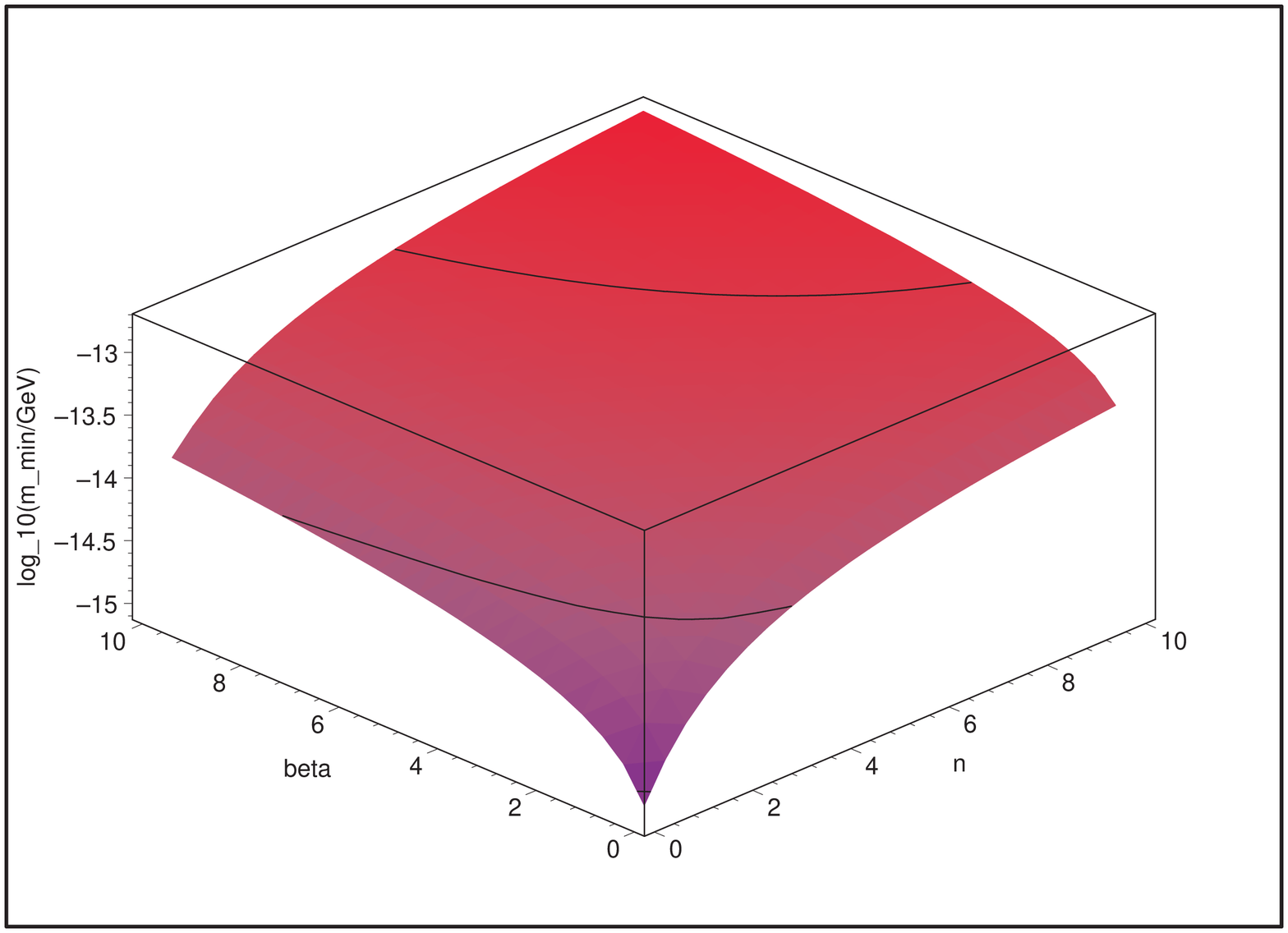}
\caption[Values of $m_\text{min}$ for the exponential potential]{The value of $\log_{10}m_\text{min}$ in Earth's atmosphere, plotted as a function of $n$ and $\beta$, for the exponential potential.  Contours are drawn for integer values of $\log_{10}m_\text{min}$.}
\label{exponential_mmin}
\end{figure}
We are now in a position to estimate the range of the chameleon force and to see why the chameleon may be obscured from terrestrial observation; we will do this more rigourously in Section \ref{copperballexperiment}, once we have laid the necessary mathematical groundwork in the intervening sections.

Below, we will assume a single matter species representing the Earth's atmosphere, with $\tilde{\rho}_\text{atm}=10^{-3}\,\text{g}/\text{cm}^3\approx4\times10^{-21}\,\text{GeV}^4$ and $w=0$.

If $\phi\gg M$, then $V_\phi$ is independent of which of the two potentials we are working with:
\begin{displaymath}
\frac{\partial}{\partial\phi}\left(\frac{M^{4+n}}{\phi^n}\right)\approx\frac{\partial}{\partial\phi}\left(M^4e^{M^n/\phi^n}\right).
\end{displaymath}
Therefore we may assume $V\left(\phi\right)=M^{4+n}\phi^{-n}$ in this subsection for the sake of mathematical simplicity.

With this assumption, we can determine $m_\text{min}$ exactly for the Earth's atmosphere.  Equation (\ref{phimin}) becomes
\begin{gather*}
-n\frac{M^{4+n}}{\phi_\text{min}^{n+1}}+\frac{\beta}{M_\text{pl}}\tilde{\rho}_\text{atm}e^{4\beta\phi_\text{min}/M_\text{pl}}=0\\
\Rightarrow \phi_\text{min}^{n+1}e^{4\beta\phi_\text{min}/M_\text{pl}}=\frac{nM^{4+n}M_\text{pl}}{\beta\tilde{\rho}_\text{atm}}\\
\Rightarrow\frac{4\beta\phi_\text{min}}{M_\text{pl}\left(n+1\right)}e^{4\beta\phi_\text{min}/M_\text{pl}\left(n+1\right)}=\frac{4\beta}{M_\text{pl}\left(n+1\right)}\left[\frac{nM^{4+n}M_\text{pl}}{\beta\tilde\rho_\text{atm}}\right]^{1/\left(n+1\right)}\\
\Rightarrow\phi_\text{min}=\frac{M_\text{pl}\left(n+1\right)}{4\beta}W\left(\frac{4\beta}{M_\text{pl}\left(n+1\right)}\left[\frac{nM^{4+n}M_\text{pl}}{\beta\tilde\rho_\text{atm}}\right]^{1/\left(n+1\right)}\right).
\end{gather*}
Then equation (\ref{mmin}) becomes
\begin{displaymath}
m_\text{min}=\left(n\left(n+1\right)\frac{M^{4+n}}{\phi_\text{min}^{n+2}}+\frac{\beta^2}{M_\text{pl}^2}\tilde\rho_\text{atm}e^{4\beta\phi_\text{min}/M_\text{pl}}\right)^{1/2}.
\end{displaymath}

We will show later that the chameleon force obeys a Yukawa potential,
\begin{displaymath}
\Phi\left(r\right)=\alpha\frac{e^{-r/\lambda}}{r},
\end{displaymath}
with $\lambda= m_\text{min}^{-1}$.  Current experimental data give an upper bound $\lambda\lesssim1\,\text{mm}$ for a ``strong'' ($\alpha\gtrsim1$) Yukawa-potential force \cite{AHN}.  This bound translates to $m_\text{min}\gtrsim10^{-13}\,\text{GeV}$.  We will now explore this bound in the $\left(n,\beta\right)$ parameter space.

In the case of the power-law potential, we have an exact solution for $m_\text{min}$ as a function of $n$ and $\beta$, which we plot in Figure \ref{powerlaw_mmin}.  Unfortunately, for this potential we have $m_\text{min}\ll10^{-13}\,\text{GeV}$ for all reasonable values of $n$ and $\beta$.

In the case of the exponential potential, we can find $m_\text{min}$ as a function of $n$ and $\beta$ subject to the previously discussed approximation with $\phi\gg M$.  We plot this in Figure \ref{exponential_mmin}, in which the top-most contour is $m_\text{min}=10^{-13}\,\text{GeV}$.  We see that there exists a region in the $n$-$\beta$ plane which satisfies the experimental bound, although it requires both $n$ and $\beta$ to be greater than unity.
\section{Chameleon cosmology}
\label{chameleoncosmology}
In this section we will follow the argument of \cite{BvdBDKW}.

Let us assume the bare potential to be of the exponential form
\begin{displaymath}
V\left(\phi\right)=M^4\exp\left(\frac{M^n}{\phi^n}\right),
\end{displaymath}
with $M=2\times10^{-3}\,\text{eV}$.
\subsection{The cosmological chameleon equation of motion}
Assume a flat, homogeneous, isotropic Universe with the metric
\begin{displaymath}
g_{\mu\nu}=\diag\left(-1,a^2,a^2,a^2\right).
\end{displaymath}
Then, assuming $\phi$ also to be homogeneous,
\begin{align*}
\nabla^2\phi&=g^{\mu\nu}\nabla_\mu\nabla_\nu\phi\\
&=g^{\mu\nu}\partial_\mu\partial_\nu\phi-g^{\mu\nu}\Gamma_{\nu\mu}^\rho\phi_{,\rho}\\
&=g^{00}\partial_0\partial_0\phi-\left(a^{-2}\Gamma_{11}^0+a^{-2}\Gamma_{22}^0+a^{-2}\Gamma_{33}^0\right)\phi_{,0}\\
&=-\ddot{\phi}-a^{-2}\left(3a\dot{a}\right)\dot{\phi}\\
&=-\left(\ddot{\phi}+3H\dot{\phi}\right),
\end{align*}
so equation (\ref{conciseeom}) becomes
\begin{equation}
\ddot{\phi}+3H\dot{\phi}=-V_{\text{eff},\phi}(\phi),
\label{cosmiceom}
\end{equation}
which is the usual result for a spatially homogeneous scalar field.
\subsection{The Friedmann equation}
Suppose the Universe is composed of $\phi$, pressure-free matter with density $\rho_\text{m}$ coupled to $\phi$ by a coupling constant $\beta$, and radiation with density $\rho_\text{r}$.

The Friedmann equation is derived from the Einstein equation $G^{\mu\nu}=8\pi G\,T^{\mu\nu}$, which can be found, as usual, by variation of the action (\ref{total_action}) with respect to the Einstein-frame metric $g_{\mu\nu}$.  But now, for matter of species $i$ conformally coupled to $\phi$, we have
\begin{align*}
T^{\mu\nu}&\equiv-\frac{2}{\sqrt{-g}}\frac{\partial\mathcal{L}_\text{m}}{\partial g_{\mu\nu}}\\
&=-\frac{2}{\sqrt{-\tilde{g}e^{-8\beta_i\phi/M_\text{pl}}}}\frac{\partial\mathcal{L}_\text{m}}{\partial\tilde{g}_{\alpha\beta}}\frac{\partial\tilde{g}_{\alpha\beta}}{\partial g_{\mu\nu}}\\
&=-\frac{2}{\sqrt{\tilde{-g}}}\frac{\partial\mathcal{L}_\text{m}}{\partial\tilde{g}_{\mu\nu}}e^{6\beta_i\phi/M_\text{pl}}\\
&=\tilde{T}^{\mu\nu}e^{6\beta_i\phi/M_\text{pl}}.
\end{align*}
Recalling equation (\ref{densityandpressure}), we know $\tilde{T}^{00}\tilde{g}_{00}=-\tilde{\rho}$.  Using this with the above equation, we have
\begin{align*}
T^{00}&=-\tilde{T}^{00}\tilde{g}_{00}\frac{e^{6\beta_i\phi/M_\text{pl}}}{\tilde{g}_{00}}\\
&=\tilde{\rho}e^{4\beta_i\phi/M_\text{pl}}\\
&=\rho e^{-3\left(1+w_i\right)\beta_i\phi/M_\text{pl}}e^{4\beta_i\phi/M_\text{pl}}\\
&=\rho e^{\left(1-3w_i\right)\beta_i\phi/M_\text{pl}}.
\end{align*}
The Friedmann equation is therefore
\begin{equation}
3H^2M_\text{pl}^2=\frac{1}{2}\dot{\phi}^2+V\left(\phi\right)+\rho_\text{m}e^{\beta\phi/M_\text{pl}}+\rho_\text{r}.
\label{Friedmann}
\end{equation}

We further make the natural definitions
\begin{displaymath}
\rho_\text{critical}\equiv\frac{1}{2}\dot{\phi}^2+V\left(\phi\right)+\rho_\text{m}e^{\beta\phi/M_\text{pl}}+\rho_\text{r}
\end{displaymath}
and
\begin{displaymath}
\Omega_\text{m}\equiv\frac{\rho_\text{m}e^{\beta\phi_\text{min}/M_\text{pl}}}{\rho_\text{critical}}
\end{displaymath}
for this cosmology, given the assumption of a spatial flatness.
\subsection{Cosmological evolution of the chameleon}
\label{cosmologicalevolution}
We know from equation (\ref{conciseeom}) that the chameleon seeks out the minimum of $V_\text{eff}$.  But that minimizing value, $\phi_\text{min}$, now changes over time, as $\rho_\text{m}$ is diluted by the expansion of the Universe.  We would like to know if this process is adiabatic; that is, does the chameleon field $\phi(t)$ keep up with the minimum $\phi_\text{min}(t)$ as the Universe expands?

To answer this question, we need to know both how fast $\phi_\text{min}(t)$ runs away from $\phi(t)$ and how fast $\phi(t)$ can chase after it.  Equation (\ref{phimin}) tells us that the characteristic time for the evolution of $\phi_\text{min}(t)$ is roughly that of the evolution of $\rho_\text{m}$, i.e.,
\begin{displaymath}
\left|\frac{\rho_\text{m}}{\dot{\rho}_\text{m}}\right|\sim H^{-1},
\end{displaymath}
where $H^{-1}$ is the Hubble time.

The characteristic response time for the field $\phi(t)$ can be found by making the damped simple harmonic oscillator approximation to equation (\ref{cosmiceom}),
\begin{equation}
\ddot{\phi}+3H\dot{\phi}=-\omega^2\left(\phi-\phi_\text{min}\right),
\label{oscillatorapproximation}
\end{equation}
where the constant $\omega$ is given by
\begin{displaymath}
\omega^2=\left.\frac{dV_{\text{eff},\phi}}{d\phi}\right|_{\phi=\phi_\text{min}}=V_{\text{eff},\phi\phi}\left(\phi_\text{min}\right)=m_\text{min}^2,
\end{displaymath}
with the last equality coming from equation (\ref{mmin}).  The characteristic frequency of the oscillator is of course $\omega=m_\text{min}$.  If the oscillator is underdamped, then its characteristic response time is the inverse of the frequency, $m_\text{min}^{-1}$.  Note that the condition for underdamping is that $2m_\text{min}>3H$.

In order for $\phi(t)$ to keep up with $\phi_\text{min}\left(t\right)$, we must therefore have $m_\text{min}^{-1}\ll H^{-1}$; that is,
\begin{displaymath}
m_\text{min}\gg H,
\end{displaymath}
which is consistent with the oscillator being underdamped.

Assuming that $\beta$ is of order unity and $n\gtrsim\frac{1}{2}$, we have that if $\phi\left(t\right)$ is slowly-rolling, then $m_\text{min}\gg H$ from the earliest times (ie.\ the end of inflation) until today (Appendix \ref{proof_mmin_H}).  If $m_\text{min}\gg H$, then $\phi_\text{min}\left(t\right)$ is slowly-rolling (Appendix \ref{proof_slowroll}).  But $m_\text{min}\gg H$ also gives $\phi\left(t\right)\approx\phi_\text{min}\left(t\right)$, and therefore $\phi\left(t\right)$ is slowly-rolling too.  This picture is consistent:  If $\phi\left(t\right)\approx\phi_\text{min}\left(t\right)$ at some initial time, then $\phi\left(t\right)\approx\phi_\text{min}\left(t\right)$ from that initial time until today.

However, in the future, dilution of $\rho_\text{m}$ and $\rho_\text{r}$ will allow $V\left(\phi\right)\approx M^4$ to dominate the energy budget.  At this stage, the Universe will be expanding exponentially as a de Sitter spacetime, so $H$ will become constant.  However, $m_\text{min}$ will decrease quickly as $\rho_\text{m}$ decreases, and eventually we will have $m\sim H$.  After this time, $\phi$ will no longer be able to keep up with $\phi_\text{min}$, and as the contribution to $V_\text{eff}\left(\phi\right)$ from the matter term $\rho e^{\beta\phi/M_\text{pl}}$ flattens out, the dynamics of $\phi$ will be governed solely by the bare potential $V\left(\phi\right)$, as in normal quintessence.
\subsection{The chameleon in the early Universe}
It remains to discuss the initial conditions mentioned above.  We now sketch this out, referring the reader to \cite{BvdBDKW} for the details.  We assume that the Universe undergoes an initial period of inflation driven by an inflaton field with $w\approx-1$ which is coupled to $\phi$ in the same manner as matter.  Then equation (\ref{veff}) gives
\begin{displaymath}
V_{\text{eff}}(\phi)\approx M^4\exp\left(\frac{M^n}{\phi^n}\right)+\rho_\text{vac}e^{4\beta_i\phi/M_\text{pl}},
\end{displaymath}
where $\rho_\text{vac}$ is the energy density of the inflaton field.  Since $\rho_\text{vac}$ is roughly constant during inflation, the potential function $V_\text{eff}\left(\phi\right)$ is time-independent, so equation (\ref{cosmiceom}) is the equation of a damped oscillator.

Combining equations (\ref{phimin}) and (\ref{mmin}) and the Friedmann equation, as we did in section \ref{cosmologicalevolution}, (or simply mapping $\beta\mapsto4\beta$ and $\Omega_\text{m}\mapsto\Omega_\text{vac}$ in equation (\ref{mminoverHequality})) gives
\begin{displaymath}
\frac{m_\text{min}^2}{H^2}\approx12\beta\Omega_\text{vac}\frac{M_\text{pl}}{\phi_\text{min}}\left[1+n+n\left(\frac{M}{\phi_\text{min}}\right)^n\right]+48\beta^2\Omega_\text{vac},
\end{displaymath}
with
\begin{displaymath}
\Omega_\text{vac}\equiv\frac{\rho_\text{vac}e^{4\beta\phi_\text{min}/M_\text{pl}}}{\rho_\text{critical}}.
\end{displaymath}
Since $\beta$ and $\Omega_\text{vac}$ are of order unity, we certainly have, for the discriminant of equation (\ref{oscillatorapproximation}),
\begin{displaymath}
\frac{4m_\text{min}^2}{\left(3H\right)^2}>\frac{4}{9}\cdot48\beta^2\Omega_\text{vac}>1,
\end{displaymath}
so that the oscillator is underdamped in the harmonic approximation.  We therefore have that $\phi$ is driven exponentially towards $\phi_\text{min}$ during inflation:
\begin{displaymath}
\left|\phi-\phi_\text{min}\right|\lesssim e^{-3Ht/2}\propto a^{-3/2}.
\end{displaymath}
Note that for the harmonic oscillator approximation to be valid, we must assume initial conditions such that $m\gtrsim H$ initially.

At the end of inflation, the inflaton decays into mostly radiation, plus some matter and small excitations of the field $\phi$.  Because $1-3w=0$ for radiation, it does not couple to $\phi$ and so $\phi_\text{min}$ increases dramatically during reheating.  We still have $m\gg H$, so $\phi$ is driven to and past $\phi_\text{min}$.  But as $\phi$ increases, $m$ decreases until friction takes over, at which point $\phi$ gets stuck at some elevated value.

As the Universe expands and cools during the radiation era, matter species decouple from the heat bath one by one.  When a species decouples, the perfect fluid approximation $T^{\mu\nu}g_{\mu\nu}\approx-\rho$ for matter becomes invalid for a time because the decoupling particles have relativistic velocities.  This provides ``kicks'' to equation (\ref{cosmiceom}) which drive $\phi$ back towards $\phi_\text{min}$.  When the radiation era ends and the matter era begins, the driving term in equation (\ref{cosmiceom}) comes to dominate the friction term, so $\phi$ converges to $\phi_\text{min}$ and then follows it, as described in section \ref{cosmologicalevolution}.
\section{Chameleons and matter}
\subsection{The chameleon force}
The interaction of the chameleon field with matter is encapsulated by the conformal coupling of equation (\ref{conformalrelation}); so is the interaction of the spacetime geometry with matter.  Since matter fields $\psi_\text{m}^{(i)}$ couple to $g_{\mu\nu}^{(i)}$ instead of to $g_{\mu\nu}$, the worldlines of free test particles (meaning particles experiencing only gravity and the chameleon force) of species $i$ are the geodesics of $g_{\mu\nu}^{(i)}$ rather than those of $g_{\mu\nu}$ (see also \cite{Damour}).\footnote{From this it is clear that the chameleon force violates the weak Equivalence Principle only if there exist two matter species with differing values of $\beta_i$.}

The geodesic equation for the worldline $x^\mu$ of a test mass of species $i$ is
\begin{equation}
\ddot{x}^\rho+\tilde{\Gamma}_{\mu\nu}^{\rho}\dot{x}^\mu\dot{x}^\nu=0,
\label{geodesic}
\end{equation}
where $\tilde{\Gamma}_{\mu\nu}^{\rho}$ are the Christoffel symbols and a dot denotes differentiation with respect to proper time $\tilde{\tau}$, both in the $\tilde{g}_{\mu\nu}$ metric.

Using
\begin{displaymath}
\tilde{g}_{\mu\nu,\sigma}=\left(\frac{2\beta_i}{M_\text{pl}}\phi_{,\sigma}g_{\mu\nu}+g_{\mu\nu,\sigma}\right)e^{2\beta_i\phi/M_\text{pl}},
\end{displaymath}
the Christoffel symbols can be determined as follows:\footnote{The derivation of this relationship in the case of a general conformal transformation is given in \cite[pp.65--6]{FM}.}
\begin{align*}
\tilde{\Gamma}_{\mu\nu}^{\rho}&=\frac{1}{2}\tilde{g}^{\sigma\rho}\left(\tilde{g}_{\sigma\nu,\mu}+\tilde{g}_{\sigma\mu,\nu}-\tilde{g}_{\mu\nu,\sigma}\right)\\
&=\frac{1}{2}e^{-2\beta_i\phi/M_\text{pl}}g^{\sigma\rho}\left(
\begin{aligned}
\frac{2\beta_i}{M_\text{pl}}\phi_{,\mu}g_{\sigma\nu}+g_{\sigma\nu,\mu}+\frac{2\beta_i}{M_\text{pl}}\phi_{,\nu}g_{\sigma\mu}\\
\vphantom{}+g_{\sigma\mu,\nu}-\frac{2\beta_i}{M_\text{pl}}\phi_{,\sigma}g_{\mu\nu}-g_{\mu\nu,\sigma}
\end{aligned}
\right)e^{2\beta_i\phi/M_\text{pl}}\\
&=\frac{1}{2}g^{\sigma\rho}\left(g_{\sigma\nu,\mu}+g_{\sigma\mu,\nu}-g_{\mu\nu,\sigma}
\right)+\frac{\beta_i}{M_\text{pl}}g^{\sigma\rho}\left(\phi_{,\mu}g_{\sigma\nu}+\phi_{,\nu}g_{\sigma\mu}-\phi_{,\sigma}g_{\mu\nu}\right)\\
&=\Gamma_{\mu\nu}^\rho+\frac{\beta_i}{M_\text{pl}}\left(\phi_{,\mu}\delta_\nu^\rho+\phi_{,\nu}\delta_\mu^\rho-g^{\sigma\rho}\phi_{,\sigma}g_{\mu\nu}\right).
\end{align*}

Substituting this into (\ref{geodesic}) gives
\begin{align*}
0&=\ddot{x}^\rho+\Gamma_{\mu\nu}^{\rho}\dot{x}^\mu\dot{x}^\nu+\frac{\beta_i}{M_\text{pl}}\left(\phi_{,\mu}\delta_\nu^\rho+\phi_{,\nu}\delta_\mu^\rho-g^{\sigma\rho}\phi_{,\sigma}g_{\mu\nu}\right)\dot{x}^\mu\dot{x}^\nu\\
&=\ddot{x}^\rho+\Gamma_{\mu\nu}^{\rho}\dot{x}^\mu\dot{x}^\nu+\frac{\beta_i}{M_\text{pl}}\left(\phi_{,\mu}\dot{x}^\mu\dot{x}^\rho+\phi_{,\nu}\dot{x}^\rho\dot{x}^\nu-g^{\sigma\rho}\phi_{,\sigma}g_{\mu\nu}\dot{x}^\mu\dot{x}^\nu\right)\\
&=\ddot{x}^\rho+\Gamma_{\mu\nu}^{\rho}\dot{x}^\mu\dot{x}^\nu+\frac{\beta_i}{M_\text{pl}}\left(2\phi_{,\mu}\dot{x}^\mu\dot{x}^\rho+g^{\sigma\rho}\phi_{,\sigma}\right).
\end{align*}
The second term in the above equation is the familiar gravitational term, while the term in $\beta_i/M_\text{pl}$ is the chameleon force.

We see that in the non-relativistic limit, a test mass $m$ of species $i$ in a static chameleon field $\phi$ experiences a force $\vec{F}_\phi$ given by
\begin{equation}
\frac{\vec{F}_\phi}{m}=-\frac{\beta_i}{M_\text{pl}}\vec{\nabla}\phi,
\label{chameleonforce}
\end{equation}
as in \cite{KW}.  Thus, $\phi$ is the potential for the chameleon force.

\subsection{Static, spherically symmetric solutions}
We wish to find time-independent solutions $\phi(\vec{x})$ for spherically symmetric matter distributions $\tilde{\rho}(r)$ of a single pressure-free matter species\footnote{Clearly, this is identical to the case of a collection of such matter species all of whose coupling constants $\beta_i$ are equal.} in the weak-field limit.

Assuming that $g_{\mu\nu}=\eta_{\mu\nu}$, the requirement of time-independence reduces equation (\ref{conciseeom}) to
\begin{displaymath}
\vec{\nabla}^2\phi=V_{\text{eff},\phi}\left(\phi\right)=V_{,\phi}\left(\phi\right)+\frac{\beta}{M_\text{pl}}\rho e^{\beta\phi/M_\text{pl}}.
\end{displaymath}
Switching to spherical polar coordinates and assuming spherical symmetry, this becomes
\begin{equation}
\begin{aligned}
\frac{d^2\phi}{dr^2}+\frac{2}{r}\frac{d\phi}{dr}&=V_{,\phi}\left(\phi(r)\right)+\frac{\beta}{M_\text{pl}}\rho(r)e^{\beta\phi(r)/M_\text{pl}}\\
&=V_{,\phi}\left(\phi(r)\right)+\frac{\beta}{M_\text{pl}}\tilde{\rho}(r)e^{4\beta\phi(r)/M_\text{pl}}.
\end{aligned}
\label{DE}
\end{equation}

Now suppose that we have a homogeneous spherical mass of radius $R$ and density $\rho_\text{c}$ sitting in a background matter distribution with density $\rho_\infty$, with $\rho_\text{c}>\rho_\infty$.  Then
\begin{displaymath}
\tilde{\rho}(r)=
\begin{cases}
\rho_\text{c} & r<R\\
\rho_\infty & r>R.
\end{cases}
\end{displaymath}
We can define
\begin{align*}
\phi_\text{c}&\equiv\left.\phi_\text{min}\right|_{\rho=\rho_\text{c}}, & m_\text{c}&\equiv m\left(\phi_\text{c}\right),\\
\phi_\infty&\equiv\left.\phi_\text{min}\right|_{\rho=\rho_\infty}, & m_\infty&\equiv m\left(\phi_\infty\right).
\end{align*}
Note that $\phi_\text{c}<\phi_\infty$, as discussed in section \ref{potentialandeffectivepotential}.

We will now discuss solutions to equation (\ref{DE}), essentially a refinement of the solution scheme given in \cite{KW}.
\subsubsection{Three classes of solution}
Inside the sphere, equation (\ref{DE}) drives $\phi$ towards $\phi_\text{c}$, while outside the sphere, it drives $\phi$ towards $\phi_\infty$.  In order to obtain an approximate solution to (\ref{DE}), we will assume that outside the sphere, the driving term $V_\text{eff}(\phi)$ can be approximated by a harmonic potential, as we discussed in section \ref{cosmologicalevolution}.  Then for $r>R$,
\begin{displaymath}
\frac{d^2\phi}{dr^2}+\frac{2}{r}\frac{d\phi}{dr}=m_\infty^2\left(\phi-\phi_\infty\right).
\end{displaymath}
The general solution to this differential equation is
\begin{equation}
\phi(r)=A\frac{e^{-m_\infty(r-R)}}{r}+B\frac{e^{m_\infty(r-R)}}{r}+\phi_\infty,
\label{harmonicsolution}
\end{equation}
for dimensionless constants $A$ and $B$.  Imposing the condition that $\phi\rightarrow\phi_\infty$ as $r\rightarrow\infty$ gives $B=0$, so we have
\begin{equation}
\phi(r)=A\frac{e^{-m_\infty(r-R)}}{r}+\phi_\infty.
\label{outsidesolution}
\end{equation}
We recognize this as a Yukawa potential.  We now see why the characteristic interaction range $\lambda=m_\text{min}^{-1}$ from section \ref{chameleoninhiding} was justified, for the resulting chameleon force (\ref{chameleonforce}) retains the exponential suppression factor $e^{-m_\infty(r-R)}$.

For $r<R$, we will consider two classes of solution based on two different approximations to (\ref{DE}).  Firstly, define $R_\text{c}$ to divide the interval $\left[0,R\right]$ into two intervals:  $\left[0,R_\text{c}\right]$ on which $\phi\sim\phi_\text{c}$, and $\left[R_\text{c},R\right]$ on which $\phi\gg\phi_\text{c}$.  It may be the case that $R_\text{c}=0$ or $R_\text{c}=R$, so that the interval $\left[0,R\right]$ remains undivided.

We will solve (\ref{DE}) in each interval, based on two different approximations:
\begin{description}
\item[Approximation 1:  $\phi\gg\phi_\text{c}$.]
In this case the harmonic approximation to $V_\text{eff}$ is not valid.  But taking a look at Figure \ref{veffplot}, we notice that for $\phi>\phi_\text{min}$, the bare potential $V$ decays quickly and the term $\rho e^{\beta\phi/M_\text{pl}}$ comes to dominate.  In particular, we have (for both the power-law potential and the exponential potential)
\begin{equation}
V_{\text{eff},\phi}\left(\phi\right)\approx\frac{\beta}{M_\text{pl}}\rho_\text{c}e^{4\beta\phi/M_\text{pl}}\approx\frac{\beta}{M_\text{pl}}\rho_\text{c},
\label{linearapproximation}
\end{equation}
since as we have shown, $\phi\ll M_\text{pl}$ in all cases today.  Equation (\ref{DE}) now takes the form
\begin{displaymath}
\frac{d^2\phi}{dr^2}+\frac{2}{r}\frac{d\phi}{dr}\approx\frac{\beta}{M_\text{pl}}\rho_\text{c},
\end{displaymath}
with the general solution
\begin{equation}
\phi\left(r\right)=\frac{\beta}{6M_\text{pl}}\rho_\text{c}r^2+\frac{C}{r}+D\phi_\text{c},
\label{inharmonicsolution}
\end{equation}
for dimensionless constants $C$ and $D$.
\item[Approximation 2:  $\phi\sim\phi_\text{c}$.]
As with equation (\ref{harmonicsolution}), we can use the harmonic approximation
\begin{equation}
V_{\text{eff},\phi}\left(\phi\right)\approx m_\text{c}^2\left(\phi-\phi_\text{c}\right),
\label{harmonicapproximation}
\end{equation}
but this time we will write the solution as
\begin{equation}
\phi(r)=E\frac{e^{-m_\text{c}r}}{r}+F\frac{e^{m_\text{c}\left(r-R_\text{c}\right)}}{r}+\phi_\text{c},
\label{otherharmonicsolution}
\end{equation}
where $E$ and $F$ are, of course, dimensionless constants.\footnote{One may wonder why we don't use hyperbolic functions here for simplicity.  The reason for the form we choose for $\phi$ is to ensure that the values of the exponents do not become too large for numerical computation.}
\end{description}

By smoothly gluing together these solutions and imposing $\frac{d\phi}{dr}\rightarrow0$ as $r\rightarrow0$ to ensure continuity of the three-dimensional solution at the origin, we obtain an approximate solution to (\ref{DE}).  There are three cases:
\begin{description}
\item[Case 1:  $R_\text{c}=R$.]
We will call this the ``low-contrast solution''.  Here we use solution (\ref{outsidesolution}) outside the sphere and (\ref{otherharmonicsolution}) inside the sphere.  In order for $\phi\left(r\right)$ itself to have a limit as $r\rightarrow0$, we must have $E=-Fe^{-m_\text{c}R}$ in (\ref{otherharmonicsolution}), giving
\begin{displaymath}
\phi(r)=F\frac{e^{m_\text{c}\left(r-R\right)}-e^{-m_\text{c}\left(r+R\right)}}{r}+\phi_\text{c}.
\end{displaymath}
It is easy to check that $\frac{d\phi}{dr}\rightarrow0$ as $r\rightarrow0$ (using L'H\^opital's Rule).
Therefore, we have the solution
\begin{displaymath}
\phi(r)=\left\{
\begin{aligned}
&F\frac{e^{m_\text{c}\left(r-R\right)}-e^{-m_\text{c}\left(r+R\right)}}{r}+\phi_\text{c} & r&<R\\
&A\frac{e^{-m_\infty(r-R)}}{r}+\phi_\infty & r&>R.
\end{aligned}
\right.
\end{displaymath}
We have also the gradient,
\begin{displaymath}
\frac{d\phi}{dr}=\left\{
\begin{aligned}
&F\frac{m_\text{c}re^{m_\text{c}\left(r-R\right)}-e^{m_\text{c}\left(r-R\right)}+m_\text{c}re^{-m_\text{c}\left(r+R\right)}+e^{-m_\text{c}\left(r+R\right)}}{r^2} & r&<R\\
&A\frac{-m_\infty re^{-m_\infty\left(r-R\right)}-e^{-m_\infty\left(r-R\right)}}{r^2} & r&>R.
\end{aligned}
\right.
\end{displaymath}
The continuity conditions
\begin{displaymath}
\lim_{r\rightarrow R^-}\phi\left(r\right)=\lim_{r\rightarrow R^+}\phi\left(r\right)\qquad\text{and}\qquad\lim_{r\rightarrow R^-}\frac{d\phi}{dr}=\lim_{r\rightarrow R^+}\frac{d\phi}{dr}
\end{displaymath}
give two linear equations for $A$ and $F$:
\begin{gather*}
F\frac{1-e^{-2m_\text{c}R}}{R}+\phi_\text{c}=A\frac{1}{R}+\phi_\infty,\\
F\left[m_\text{c}R-1+m_\text{c}Re^{-2m_\text{c}R}+e^{-2m_\text{c}R}\right]=A\left[-m_\infty R-1\right].
\end{gather*}
The solution to these equations is
\begin{gather*}
A=\frac{\phi_\infty-\phi_\text{c}}{m_\text{c}+m_\infty+m_\text{c}e^{-2m_\text{c}R}-m_\infty e^{-2m_\text{c}R}}\left(1-m_\text{c}R-e^{-2m_\text{c}R}-m_\text{c}Re^{-2m_\text{c}R}\right),\\
F=\frac{\phi_\infty-\phi_\text{c}}{m_\text{c}+m_\infty+m_\text{c}e^{-2m_\text{c}R}-m_\infty e^{-2m_\text{c}R}}\left(1+m_\infty R\right).
\end{gather*}
\item[Case 2:  $R_\text{c}=0$.]
We will call this the ``thick-shell solution'', in keeping with \cite{KW}.  We use solution (\ref{harmonicsolution}) outside the sphere and (\ref{inharmonicsolution}) inside the sphere.  The continuity condition at the origin requires simply $C=0$ in equation (\ref{inharmonicsolution}), so we have
\begin{displaymath}
\phi(r)=\left\{
\begin{aligned}
&\frac{\beta}{6M_\text{pl}}\rho_\text{c}r^2+D\phi_\text{c} & r&<R\\
&A\frac{e^{-m_\infty(r-R)}}{r}+\phi_\infty & r&>R,
\end{aligned}
\right.
\end{displaymath}
and
\begin{displaymath}
\frac{d\phi}{dr}=\left\{
\begin{aligned}
&\frac{\beta}{3M_\text{pl}}\rho_\text{c}r & r&<R\\
&A\frac{-m_\infty re^{-m_\infty\left(r-R\right)}-e^{-m_\infty\left(r-R\right)}}{r^2} & r&>R.
\end{aligned}
\right.
\end{displaymath}
The continuity conditions give equations for $A$ and $D$:
\begin{gather*}
\frac{\beta}{6M_\text{pl}}\rho_\text{c}R^2+D\phi_\text{c}=A\frac{1}{R}+\phi_\infty\\
\frac{\beta}{3M_\text{pl}}\rho_\text{c}R=A\frac{-m_\infty R-1}{R^2},
\end{gather*}
with solutions
\begin{gather*}
A=-\frac{\beta}{3M_\text{pl}}\rho_\text{c}\frac{R^3}{1+m_\infty R},\\
D=\frac{\phi_\infty}{\phi_\text{c}}-\left(\frac{1}{1+m_\infty R}+\frac{1}{2}\right)\frac{\beta\rho_\text{c}R^2}{3\phi_\text{c}M_\text{pl}}.
\end{gather*}
\item[Case 3:  $0<R_\text{c}<R$.]
We will call this the ``thin-shell solution'', in keeping with \cite{KW}.

Now the solution is divided between three regions.  As in the case $R_\text{c}=R$, we have $E=-Fe^{-m_\text{c}R_\text{c}}$.  Then $\phi$ and $d\phi/dr$ are given by:
\begin{gather*}
\phi(r)=\left\{
\begin{aligned}
&F\frac{e^{m_\text{c}\left(r-R_\text{c}\right)}-e^{-m_\text{c}\left(r+R_\text{c}\right)}}{r}+\phi_\text{c} & r&\in\left(0,R_\text{c}\right)\\
&\frac{\beta}{6M_\text{pl}}\rho_\text{c}r^2+\frac{C}{r}+D\phi_\text{c} & r&\in\left(R_\text{c},R\right)\\
&A\frac{e^{-m_\infty(r-R)}}{r}+\phi_\infty & r&\in\left(R,\infty\right),
\end{aligned}
\right.\\
\frac{d\phi}{dr}=\left\{
\begin{aligned}
&F\frac{m_\text{c}re^{m_\text{c}\left(r-R_\text{c}\right)}-e^{m_\text{c}\left(r-R_\text{c}\right)}+m_\text{c}re^{-m_\text{c}\left(r+R_\text{c}\right)}+e^{-m_\text{c}\left(r+R_\text{c}\right)}}{r^2} & r&\in\left(0,R_\text{c}\right)\\
&\frac{\beta}{3M_\text{pl}}\rho_\text{c}r-\frac{C}{r^2} & r&\in\left(R_\text{c},R\right)\\
&A\frac{-m_\infty re^{-m_\infty\left(r-R\right)}-e^{-m_\infty\left(r-R\right)}}{r^2} & r&\in\left(R,\infty\right).
\end{aligned}
\right.
\end{gather*}
In this case there are four continuity equations:  two at $R_\text{c}$,
\begin{equation}
\begin{gathered}
F\frac{1-e^{-2m_\text{c}R_\text{c}}}{R_\text{c}}+\phi_\text{c}=\frac{\beta}{6M_\text{pl}}\rho_\text{c}R_\text{c}^2+\frac{C}{R_\text{c}}+D\phi_\text{c},\\
F\frac{m_\text{c}R_\text{c}-1+m_\text{c}R_\text{c}e^{-2m_\text{c}R_\text{c}}+e^{-2m_\text{c}R_\text{c}}}{R_\text{c}^2}=\frac{\beta}{3M_\text{pl}}\rho_\text{c}R_\text{c}-\frac{C}{R_\text{c}^2};
\end{gathered}
\label{firsttwocontinuityequations}
\end{equation}
and two at $R$,
\begin{equation}
\frac{\beta}{6M_\text{pl}}\rho_\text{c}R^2+\frac{C}{R}+D\phi_\text{c}=A\frac{1}{R}+\phi_\infty,\\
\label{thirdcontinuityequation}
\end{equation}
\begin{equation}
\frac{\beta}{3M_\text{pl}}\rho_\text{c}R-\frac{C}{R^2}=A\frac{-m_\infty R-1}{R^2}.
\label{fourthcontinuityequation}
\end{equation}
The solution to this system is quite unwieldy but is easy to obtain with Maple.  However, we still need one more equation to tell us what $R_\text{c}$ is!
\end{description}
\subsubsection{Determining \texorpdfstring{$R_\text{c}$}{R\_c}}
We wish to define $R_\text{c}\in\left[0,R\right]$ so that of the two approximations (\ref{linearapproximation}) and (\ref{harmonicapproximation}), the harmonic approximation (\ref{harmonicapproximation}) is ``better'' for $r\in\left(0,R_\text{c}\right)$ and the linear approximation (\ref{linearapproximation}) is ``better'' for $r\in\left(R_\text{c},R\right)$.

How do we define ``better''?  As $\phi$ increases from $\phi_\text{c}$, the harmonic approximation $m_\text{c}^2\left(\phi-\phi_\text{c}\right)$ increases without bound from 0.  At some point we have $m_\text{c}^2\left(\phi-\phi_\text{c}\right)=\frac{\beta}{M_\text{pl}}\rho_\text{c}$.  But we know that
\begin{displaymath}
V_{\text{eff},\phi}\approx V_\phi+\frac{\beta}{M_\text{pl}}\rho_\text{c}<\frac{\beta}{M_\text{pl}}\rho_\text{c}
\end{displaymath}
because $V_\phi<0$.  Hence, while the harmonic (Taylor) approximation is the better approximation for $m_\text{c}^2\left(\phi-\phi_\text{c}\right)<\frac{\beta}{M_\text{pl}}\rho_\text{c}$, once $m_\text{c}^2\left(\phi-\phi_\text{c}\right)>\frac{\beta}{M_\text{pl}}\rho_\text{c}$ we should switch to the linear approximation.

Therefore, $R_\text{c}$ may be defined by the following procedure:
\begin{itemize}
\item Try $R_\text{c}=R$ and compute the low-contrast solution $\phi\left(r\right)$.  If
\begin{displaymath}
m_\text{c}^2\left(\phi\left(R\right)-\phi_\text{c}\right)<\frac{\beta}{M_\text{pl}}\rho_\text{c},
\end{displaymath}
then the solution is valid and $R_\text{c}=R$.
\item Otherwise, try $R_\text{c}=0$ and compute the thick-shell solution $\phi\left(r\right)$.  If
\begin{equation}
\label{thick_shell_condition}
m_\text{c}^2\left(\phi\left(0\right)-\phi_\text{c}\right)>\frac{\beta}{M_\text{pl}}\rho_\text{c},
\end{equation}
then the solution is valid and $R_\text{c}=0$.
\item Otherwise, $R_\text{c}$ is defined by the equation
\begin{displaymath}
m_\text{c}^2\left(\phi\left(R_\text{c}\right)-\phi_\text{c}\right)=\frac{\beta}{M_\text{pl}}\rho_\text{c},
\end{displaymath}
and $\phi\left(r\right)$ is given by the thin-shell solution.  Substituting the $r\rightarrow R_\text{c}^-$ limit of the thin-shell solution, this becomes
\begin{gather}
m_\text{c}^2F\frac{1-e^{-2m_\text{c}R_\text{c}}}{R_\text{c}}=\frac{\beta}{M_\text{pl}}\rho_\text{c}\notag\\
\Rightarrow F=\frac{\beta\rho_\text{c}R_\text{c}}{m_\text{c}^2M_\text{pl}\left(1-e^{-2m_\text{c}R_\text{c}}\right)}.
\label{R0equation}
\end{gather}
The solution to the four continuity equations gives $F$ in terms of known quantities and $R_\text{c}$, so upon substituting that solution into (\ref{R0equation}), we have a single equation which determines $R_\text{c}$ in terms of known quantities but which must be solved numerically.
\end{itemize}
\subsubsection{Automating the solution}
The algorithm described above is straightforward to follow and has been encoded as a Maple script (Listing \ref{sphereprofile}).  This allows us to calculate the effect of the chameleon field for any spherically symmetric two-zone density model.

\begin{figure}[ht]
\centering
\includegraphics[scale=0.5,angle=270]{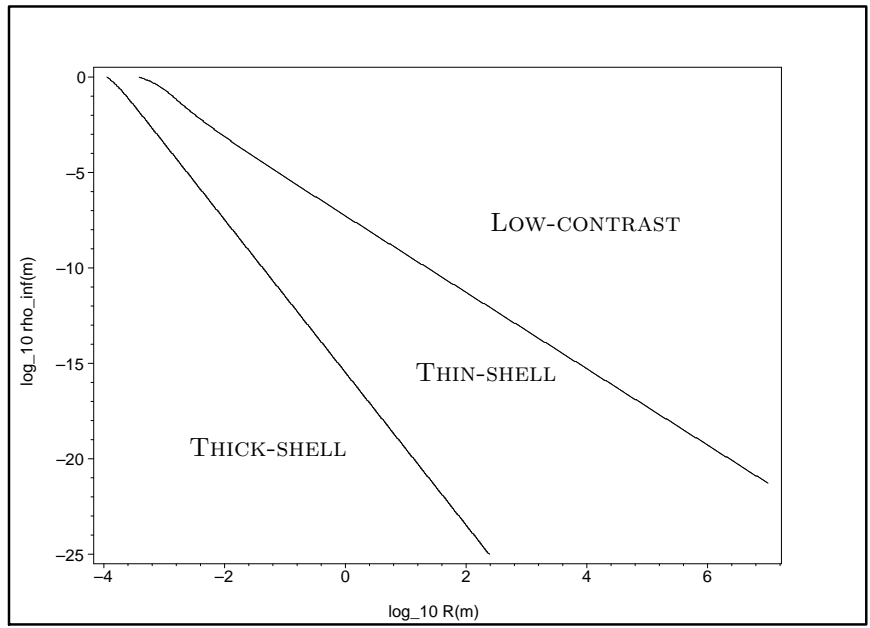}
\put(-5,-3){\textsc{Low-contrast}}
\put(-6,-5){\textsc{Thin-shell}}
\put(-9,-6){\textsc{Thick-shell}}
\caption[Solution type as a function of $R$ and $\rho_\infty$]{The regions on which the thick-shell, thin-shell and low-contrast solutions hold, as a function of $R$ and $\rho_\infty$.  The exponential potential is assumed, with $n=1$, $\beta=1$ and $\rho_\text{c}=10\,\text{g}\,\text{cm}^{-3}$.}
\label{solid_contour}
\end{figure}

The algorithm can also be used to plot the dependence of the solution type on parameters like $R$ and $\rho_\infty$, as we have done in Figure \ref{solid_contour}.

\subsubsection{The thin-shell suppression factor}
Of greatest interest to us is the behaviour of the field $\phi$ outside the sphere, for it is there that experiments to measure the chameleon force must take place.

Here we will reconnect with the previous literature by showing that the thick-shell and thin-shell solutions agree with \cite{KW} in the limit $m_\infty R\ll1$ and (in the thin-shell case) $F=0$, $\left(R-R_\text{c}\right)/R\ll1$.  Note that the low-contrast solution, while not present explicitly in Khoury and Weltman's work, is simply the $R_\text{c}\rightarrow R$ limit of the thin-shell solution.

In all cases, the exterior approximate solution is
\begin{displaymath}
\phi\left(r\right)=A\frac{e^{-m_\infty(r-R)}}{r}+\phi_\infty.
\end{displaymath}
The dimensionless constant $A$ tells us about the magnitude of $\phi$ and thus of the chameleon force.

In the thick-shell case, assuming $m_\infty R\ll1$, we have
\begin{align*}
A&=-\frac{\beta}{3M_\text{pl}}\rho_\text{c}\frac{R^3}{1+m_\infty R}\\
&\approx-\frac{\beta}{4\pi M_\text{pl}}\left(\frac{4}{3}\pi R^3\rho_\text{c}\right).
\end{align*}

In the thin-shell case, Khoury and Weltman assume $\phi=\phi_\text{c}$ for $r<R_\text{c}$, which translates for us to $F=0$.  Then the two continuity equations (\ref{firsttwocontinuityequations}) give
\begin{gather*}
C=\frac{\beta}{3M_\text{pl}}\rho_\text{c}R_\text{c}^3,\\
D=1-\frac{\beta\rho_\text{c}R_\text{c}^2}{2M_\text{pl}}\frac{1}{\phi_\text{c}}.
\end{gather*}
On substituting into the continuity equation (\ref{fourthcontinuityequation}), we get
\begin{gather}
\frac{\beta}{3M_\text{pl}}\rho_\text{c}R-\frac{\beta}{3M_\text{pl}}\rho_\text{c}\frac{R_\text{c}^3}{R^2}\approx-\frac{A}{R^2}\notag\\
\Rightarrow A\approx-\frac{\beta}{3M_\text{pl}}\rho_\text{c}\left(R^3-R_\text{c}^3\right).
\label{Aapproximation}
\end{gather}
Next, substituting into (\ref{thirdcontinuityequation}) gives
\begin{gather*}
\frac{\beta}{6M_\text{pl}}\rho_\text{c}R^2+\frac{\beta}{3M_\text{pl}}\rho_\text{c}\frac{R_\text{c}^3}{R}+\phi_\text{c}-\frac{\beta\rho_\text{c}R_\text{c}^2}{2M_\text{pl}}=-\frac{\beta}{3M_\text{pl}}\rho_\text{c}R^2+\frac{\beta}{3M_\text{pl}}\rho_\text{c}\frac{R_\text{c}^3}{R}+\phi_\infty\\
\Rightarrow\frac{\beta}{2M_\text{pl}}\rho_\text{c}R^2+\phi_\text{c}-\frac{\beta\rho_\text{c}R_\text{c}^2}{2M_\text{pl}}=\phi_\infty\\
\Rightarrow R^2-R_\text{c}^2=\frac{2M_\text{pl}}{\beta\rho_\text{c}}\left(\phi_\infty-\phi_\text{c}\right).
\end{gather*}
Finally, Taylor-expand equation (\ref{Aapproximation}) in $R_\text{c}$ about $R$ to get
\begin{align*}
A&\approx-\frac{\beta}{3M_\text{pl}}\rho_\text{c}\frac{3}{2}R\left(R^2-R_\text{c}^2\right)\\
&=-\frac{\beta}{3M_\text{pl}}\rho_\text{c}\frac{3}{2}R\frac{2M_\text{pl}}{\beta\rho_\text{c}}\left(\phi_\infty-\phi_\text{c}\right)\\
&=-\frac{\beta}{4\pi M_\text{pl}}\left(\frac{4}{3}\pi R^3\rho_\text{c}\right)\frac{3M_\text{pl}\left(\phi_\infty-\phi_\text{c}\right)}{\beta\rho_\text{c}R^2}.
\end{align*}
The approximate external solutions may thus be written in the following form (as they appear in \cite{KW}):
\begin{align*}
\phi_\text{thick}\left(r\right)&\approx-\frac{\beta}{4\pi M_\text{pl}}\left(\frac{4}{3}\pi R^3\rho_\text{c}\right)\frac{e^{-m_\infty(r-R)}}{r}+\phi_\infty\\
\phi_\text{thin}\left(r\right)&\approx-\frac{\beta}{4\pi M_\text{pl}}\left(\frac{4}{3}\pi R^3\rho_\text{c}\right)\left(3\frac{M_\text{pl}\left(\phi_\infty-\phi_\text{c}\right)}{\beta\rho_\text{c}R^2}\right)\frac{e^{-m_\infty(r-R)}}{r}+\phi_\infty
\end{align*}

The difference between the thin-shell external solution and the thick-shell external solution is the ``thin-shell suppression factor'' \cite[equation 16]{KW} $3\Delta R/R$, where
\begin{displaymath}
\frac{\Delta R}{R}\equiv\frac{M_\text{pl}\left(\phi_\infty-\phi_\text{c}\right)}{\beta\rho_\text{c}R^2}.
\end{displaymath}
Khoury and Weltman use this factor to distinguish between the thick-shell case ($\Delta R\gtrsim R$) and the thin-shell case ($\Delta R\ll R$).  Our criterion (\ref{thick_shell_condition}) for the thick-shell condition can be translated in terms of $\Delta R/R$ as follows:
\begin{gather*}
m_\text{c}^2\left(\phi(0)-\phi_\text{c}\right)>\frac{\beta}{M_\text{pl}}\rho_\text{c}\\
\Leftrightarrow m_\text{c}^2\left(\phi_\infty-\phi_\text{c}-\frac{\beta\rho_\text{c}R^2}{2M_\text{pl}}\right)>\frac{\beta\rho_\text{c}}{M_\text{pl}}\\
\Leftrightarrow\frac{M_\text{pl}\left(\phi_\infty-\phi_\text{c}\right)}{\beta\rho_\text{c}R^2}-\frac{1}{2}>\frac{1}{m_\text{c}^2R^2}\\
\Leftrightarrow\frac{\Delta R}{R}>\frac{1}{m_\text{c}^2R^2}+\frac{1}{2}.
\end{gather*}

We may, in general, define a ``chameleon suppression factor''
\begin{align*}
W&\equiv -A\left[\frac{\beta}{4\pi M_\text{pl}}\left(\frac{4}{3}\pi R^3\rho_\text{c}\right)\right]^{-1}\\
&=-A\frac{3M_\text{pl}}{\beta R^3\rho_\text{c}}.
\end{align*}
Clearly we have $W\approx1$ in the thick-shell case and $0<W<1$ in the thin-shell and low-contrast cases.  This gives us a means of quantifying the ``thickness of shell'' of a object in a given background density.
\subsubsection{Generalizations}
There are two important generalizations of our solution to equation (\ref{DE}).

The first is to situations in which $\rho_\text{c}<\rho_\infty$.  This is the case for modelling the chameleon field in a vacuum chamber, for example, to show that the chameleon field cannot be detected in laboratory vacuum experiments.  To adapt our solution to this case, we could change the solution regions so that $R_\text{c}\geq R$.  That is, we would use the harmonic approximation inside the sphere, the linear approximation on the interval $\left(R,R_\text{c}\right)$, and the harmonic approximation again on the interval $\left(R_\text{c},\infty\right)$.  However, a vacuum-chamber model has already been provided in \cite{KW}, which shows that if the radius of the vacuum chamber is $R_\text{vac}$, then the mass of the chameleon field inside the chamber is $R_\text{vac}^{-1}$.  This will allow us in section \ref{copperballexperiment} to use our existing solution algorithm with $\rho_\infty$ tuned so that $m_\text{vac}=R_\text{vac}^{-1}$.

The second generalization is to the case of more than two density regions, for example in modelling the Earth with its atmosphere.  We can adapt our solution to this case by writing the density regions as a sequence of intervals $\left(R_i,R_{\text{c},i}\right)\cup\left(R_{\text{c},i},R_{i+1}\right)$ and solving a system of equations for $R_{\text{c},i}$ and for the continuity conditions at all the interval boundaries.  However, if all adjacent pairs of intermediate density regions satisfy the thin-shell condition, the solution for the whole system is simply the piecewise connection of the sequence of two-region solutions.
\section{Detecting the chameleon field}
Reference \cite{KW} provides a discussion of several experimental models which demonstrates why the chameleon, if it exists, is capable of evading experimental detection to date.  We will provide examples here, assuming for concreteness an exponential potential for $\phi$ with $\beta=1$ and $n=1$.
\subsection{The chameleon field for the Earth}
\begin{figure}[ht]
\centering
\includegraphics[scale=0.43,angle=270]{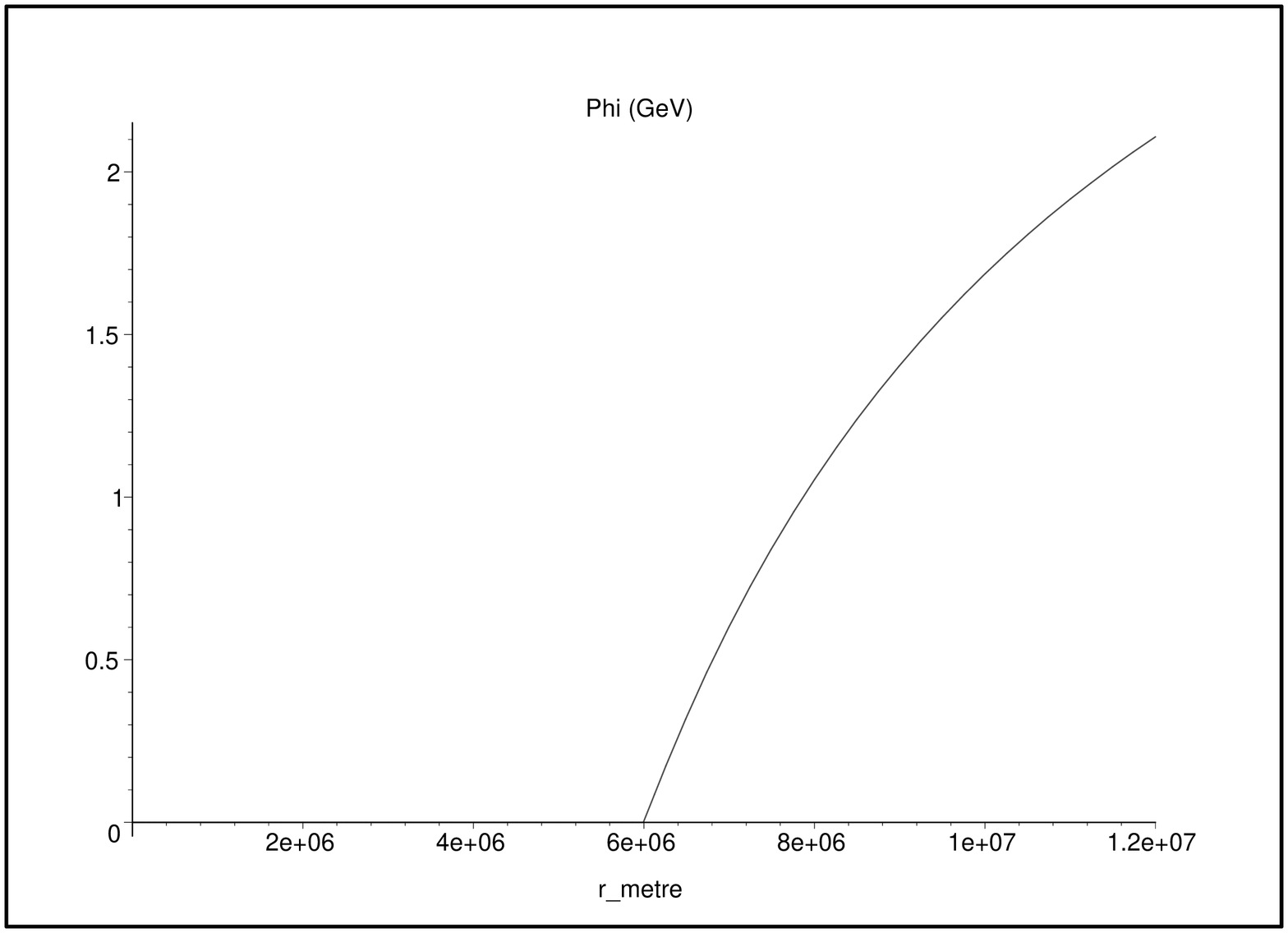}
\includegraphics[scale=0.43,angle=270]{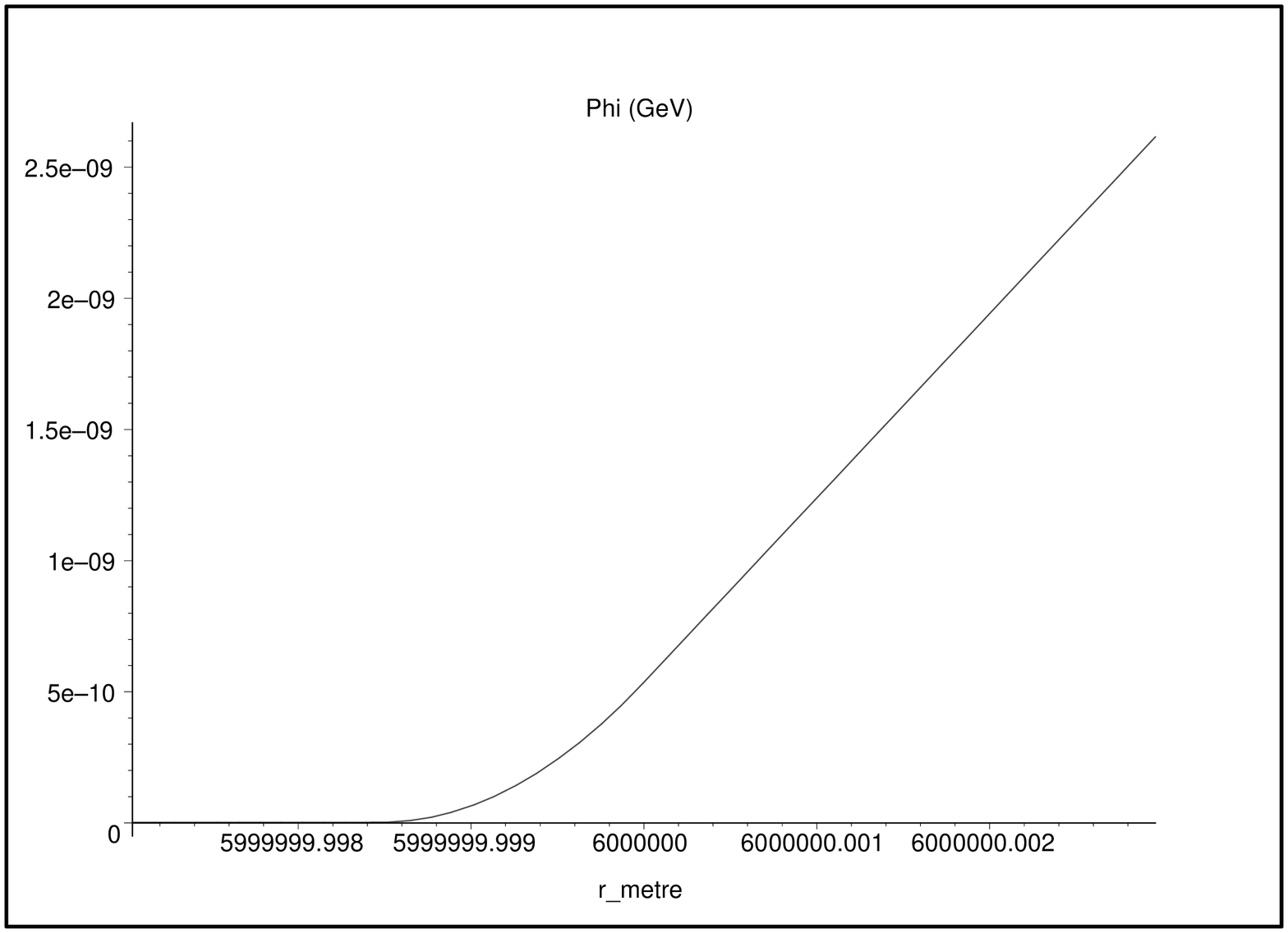}
\caption[The Earth's chameleon field]{The Earth's chameleon field (above) and a close-up of the region $\left[R_\text{c},R+\left(R-R_\text{c}\right)\right]$ (below).}
\label{earthphi}
\end{figure}
\begin{figure}[ht]
\centering
\includegraphics[scale=0.43,angle=270]{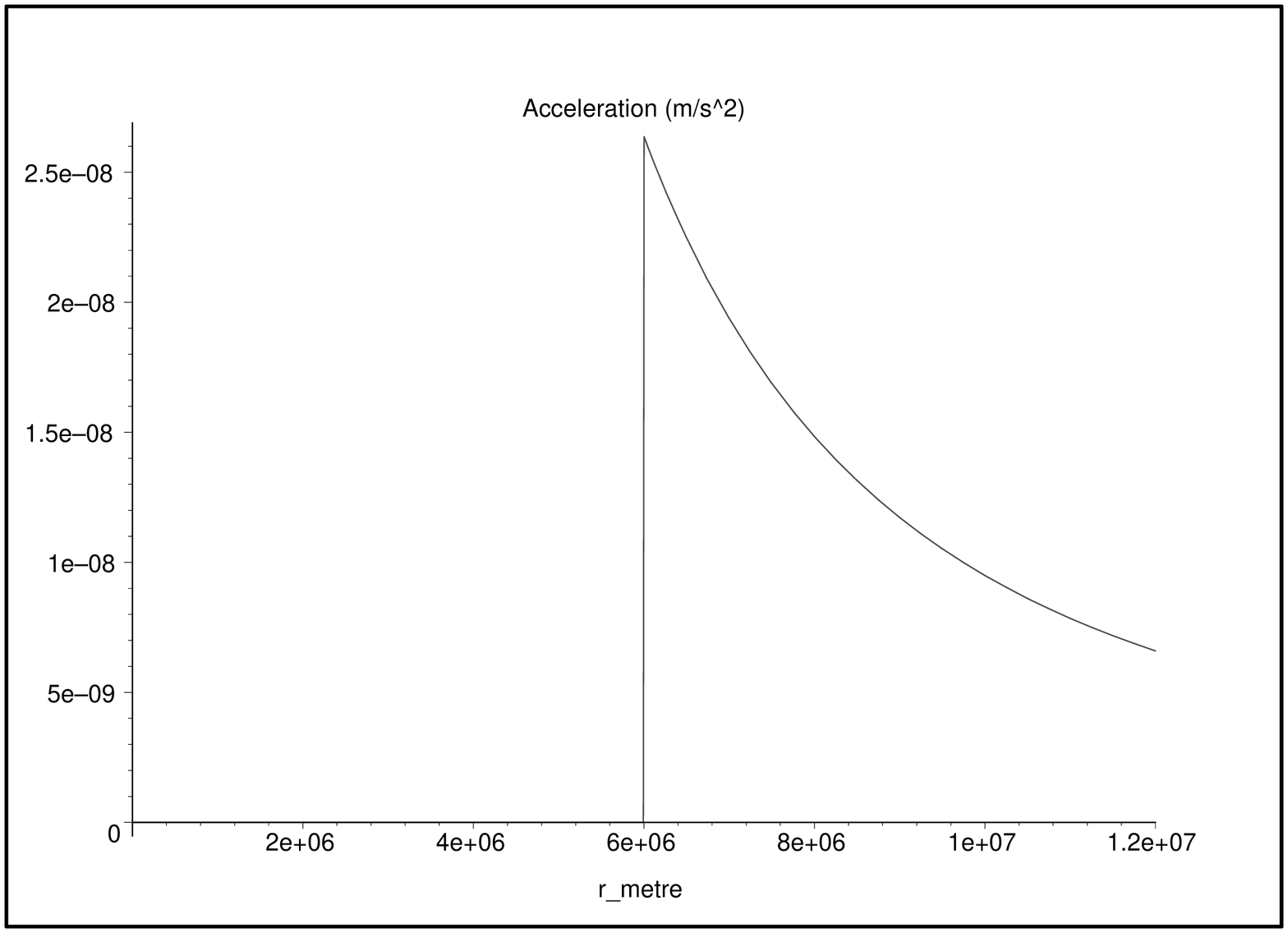}
\includegraphics[scale=0.43,angle=270]{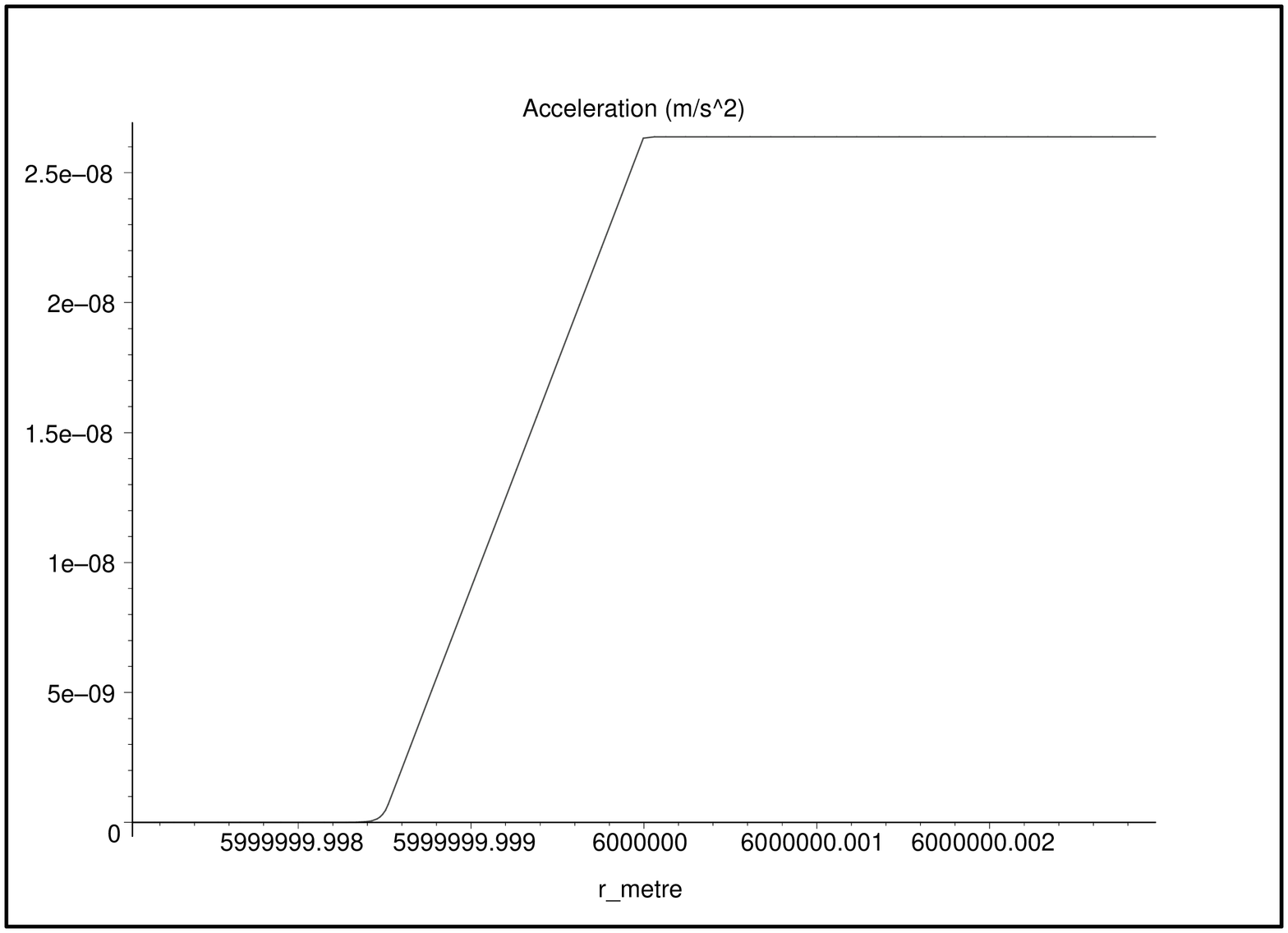}
\caption[Acceleration of test masses due to the Earth's chameleon field]{Acceleration of test masses due to the Earth's chameleon field.  The length scales match those in Figure \ref{earthphi}.}
\label{earthaccel}
\end{figure}
Let us model the Earth as a sphere of radius $6\times10^6\,\text{m}$ and density $10\,\text{g}\,\text{cm}^{-3}$ surrounded by an interplanetary medium of density $10^{-24}\,\text{g}\,\text{cm}^{-3}$ \cite{Eddington}.  Ignoring the atmosphere,\footnote{The same approximate result is obtained in \cite{KW} including the atmosphere.} our algorithm gives a thin-shell solution (Figures \ref{earthphi} and \ref{earthaccel}) with
\begin{gather*}
A\approx-10^{23},\\
W\approx10^{-9}.
\end{gather*}
Thus, deviations from Newtonian gravity due to the chameleon field of the Earth are suppressed by nine orders of magnitude by the thin-shell effect.
\subsection{The chameleon field for a small copper ball}
\label{copperballexperiment}
Now consider a copper ball of radius $1\text{cm}$ and density $\rho_\text{c}=8.92\,\text{g}\,\text{cm}^{-3}$, such as those used in small-scale gravity experiments like \cite{HNSS}.

\begin{figure}[ht]
\centering
\includegraphics[scale=0.43,angle=270]{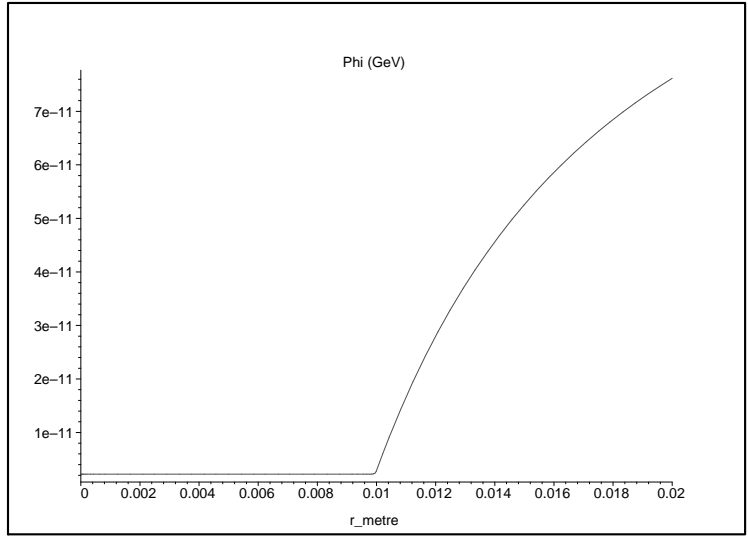}
\caption[The chameleon field of a copper ball in air]{The chameleon field of a copper ball in air.}
\label{Cuballatmos}
\end{figure}
In the atmosphere, with density $\rho_\infty=1.3\times10^{-3}\,\text{g}\,\text{cm}^{-3}$, our algorithm gives a low-contrast solution (Figure \ref{Cuballatmos}) with
\begin{gather*}
A\approx-10^4,\\
W\approx0.01.
\end{gather*}

\begin{figure}[ht]
\centering
\includegraphics[scale=0.43,angle=270]{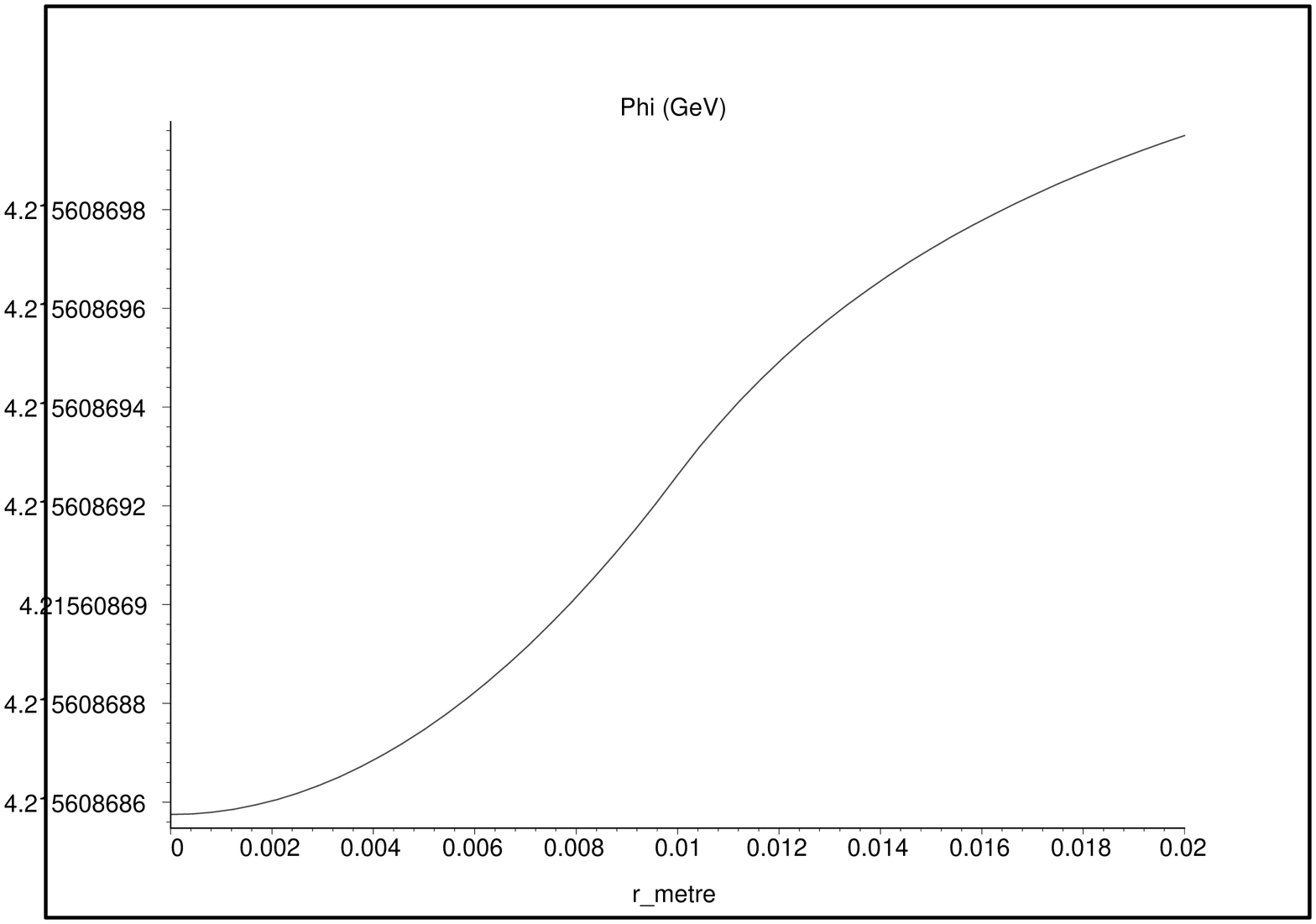}
\caption[The chameleon field of a copper ball in space]{The chameleon field of a copper ball in space.}
\label{Cuballspace}
\end{figure}
In space, with density $\rho_\infty=10^{-24}\,\text{g}\,\text{cm}^{-3}$, our algorithm gives a thick-shell solution (Figure \ref{Cuballspace}) with
\begin{gather*}
A\approx-10^6,\\
W\approx1.
\end{gather*}

\begin{figure}[ht]
\centering
\includegraphics[scale=0.43,angle=270]{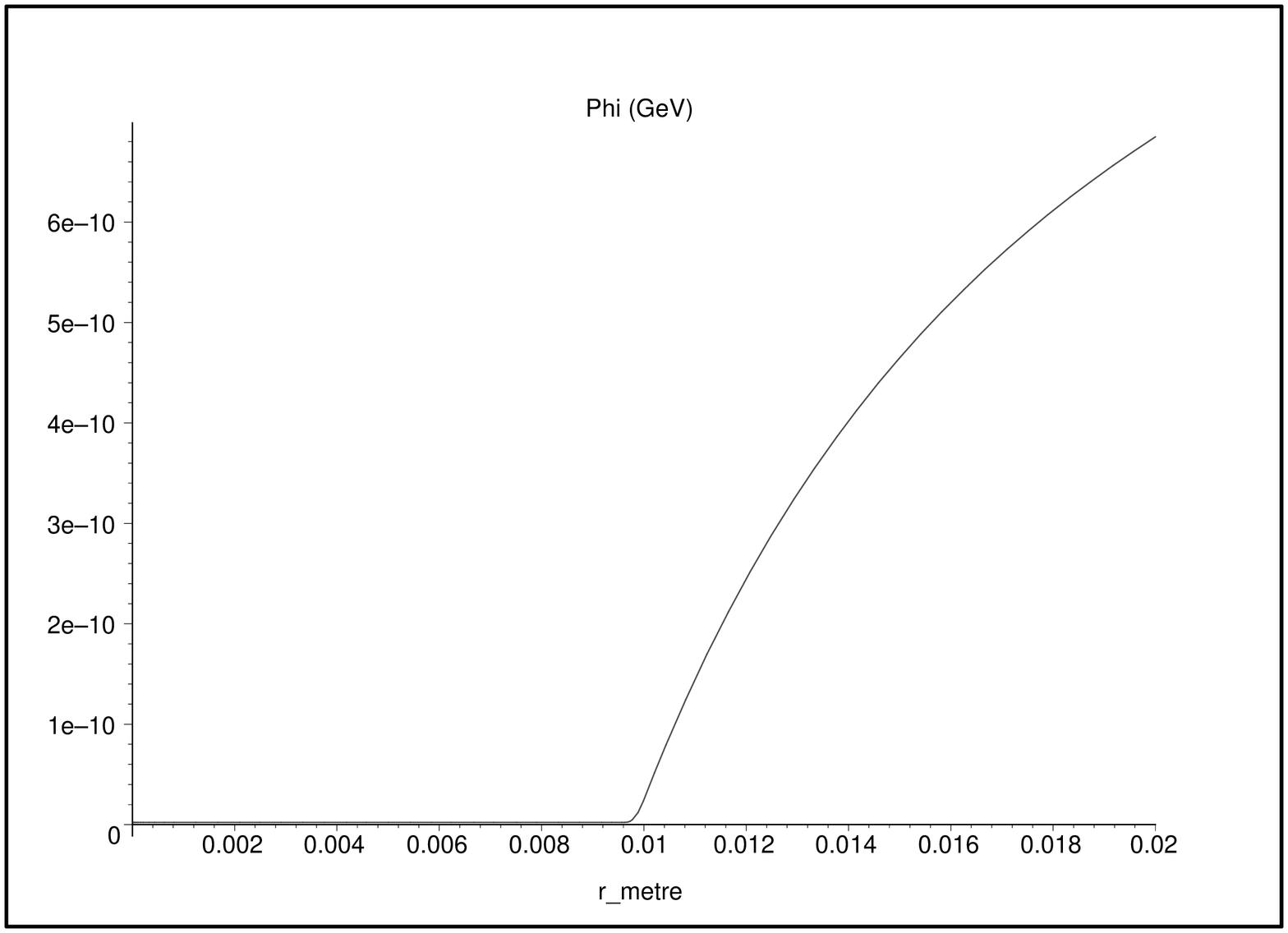}
\caption[The chameleon field of a copper ball in a vacuum chamber]{The chameleon field of a copper ball in a vacuum chamber.}
\label{Cuballvacuumchamber}
\end{figure}
Finally, we consider the chameleon profile for such a ball inside a vacuum chamber of radius $R_\text{vac}$, as in \cite{HNSS}.  As shown in \cite{KW}, the chameleon field has a mass $m_\text{vac}\approx R_\text{vac}^{-1}$ inside the chamber.  Using equations (\ref{phimin}) and (\ref{mmin}), we can assign an effective density $\rho_\infty\left(R_\text{vac}\right)$ for the vacuum chamber and then apply our solution algorithm to find the chameleon profile for the copper ball within the chamber.  A convenient example is $\rho_\infty=10^{-5}\,\text{g}\,\text{cm}^{-3}$, corresponding to $m_\text{vac}\approx1.6\times10^{-16}\,\text{GeV}$ and $R_\text{vac}\approx1.2\,\text{m}$.  Then our algorithm yields a thin-shell solution (Figure \ref{Cuballvacuumchamber}) with
\begin{gather*}
A\approx-10^5,\\
W\approx0.1.
\end{gather*}

The coupling strength of the Yukawa-potential interaction between two identical masses is (cf.\ \cite{KW})
\begin{displaymath}
\alpha=2\beta^2W^2.
\end{displaymath}
We see that for $W<0.1$, the experimental constraints of Hoskins et al.\ \cite[Figure 13]{HNSS} are satisfied for $\beta\lesssim1$.  Thus, the chameleon field can be hidden from their experiment.

\begin{figure}[ht]
\centering
\includegraphics[scale=0.43,angle=270]{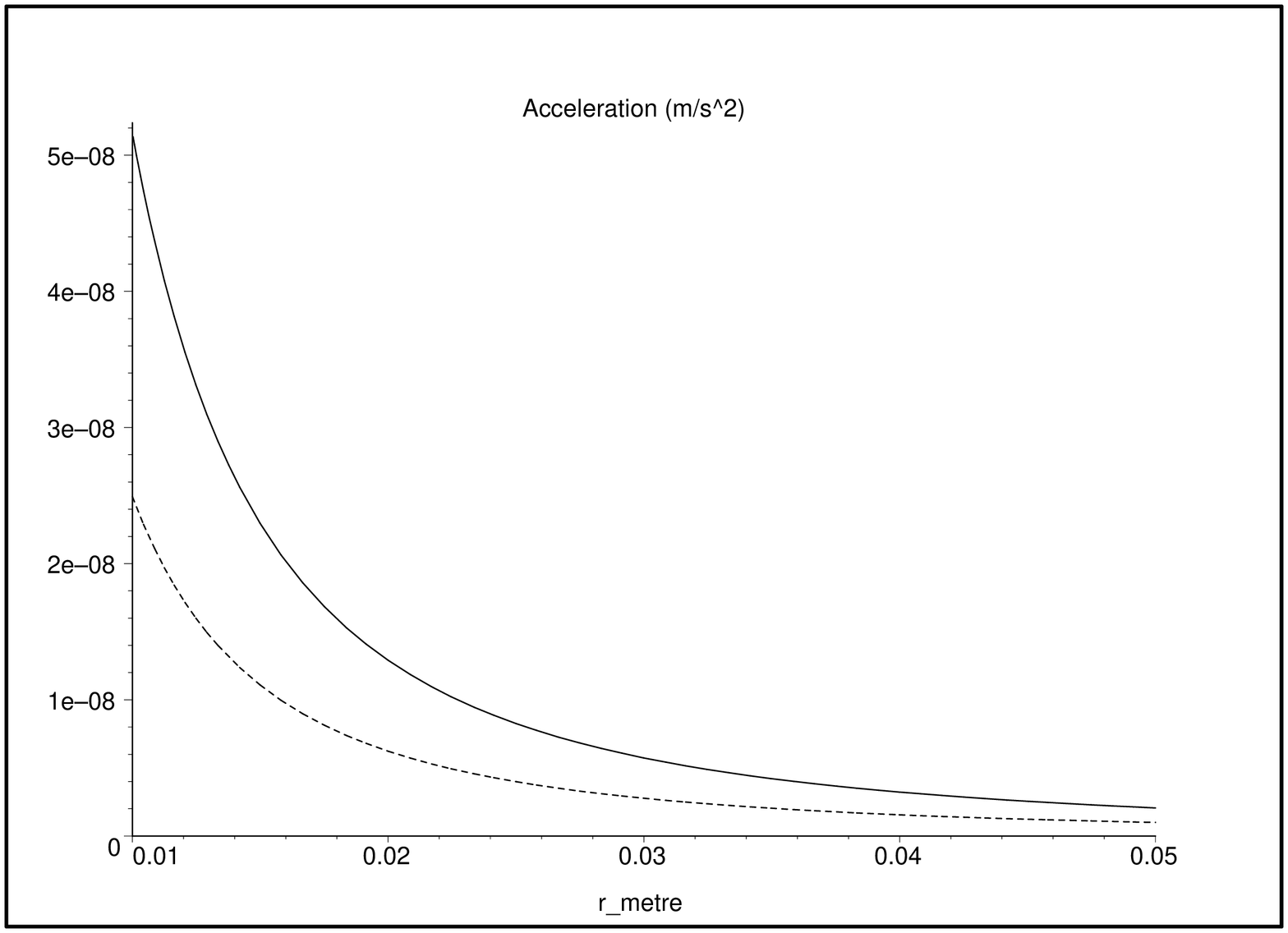}
\caption[The acceleration of a test mass near a copper ball in space]{The acceleration of a test mass near a copper ball in space due to chameleon attraction (solid line) and due to Newtonian gravity (dashed line).}
\label{Cuballspacewithgravity}
\end{figure}
However, if the experiment were repeated in orbit with $W\approx1$, the chameleon field of the copper ball would introduce a correction of order unity to its gravitational field:  The Newtonian gravitational acceleration of a test mass $5\,\text{cm}$ from the centre of the copper ball is
\begin{displaymath}
G\frac{M_\text{ball}}{\left(2\,\text{cm}\right)^2}\approx G\frac{37\,\text{g}}{4\,\text{cm}^2}\approx 1.0\times10^{-9}\frac{\text{m}}{\text{s}^2},
\end{displaymath}
and the chameleon acceleration of such a test mass is $2.0\times10^{-9}\,\text{m}\,\text{s}^{-2}$.  See Figure \ref{Cuballspacewithgravity} for a comparison of these forces at short range.
\subsection{Satellite detection of the chameleon field}
The SEE Project \cite{SEE1, SEE2} proposed to test gravity in space, including searching for a fifth force between two test bodies.  Supposing that the SEE satellite is a sphere of radius $2\,\text{m}$ and density $6\text{g}\,\text{cm}^{-3}$, it satisfies the thick-shell condition for its chameleon field.\footnote{Of course, satellites have thick shells for a large range of radii and densities.}  Thus, $\phi\approx\phi_\infty$ inside the satellite,\footnote{Actually, since the experiment will take place in orbit at an altitude of about $1000\,\text{km}$, $\phi_\infty$ will itself be damped by the presence of the Earth, but the qualitative behaviour of $\phi\left(r\right)$ near the test mass will be unaffected.} so an experiment taking place within the satellite is effectively taking place in space.  As we saw above, test masses in space should exhibit a mutual attraction due to the chameleon force of the same order as their gravitational attraction, assuming $\beta\sim1$.  The SEE Project, or another space experiment like it, such as APSIS \cite{APSIS}, would thus provide a test for the existence of a chameleon field.

The satellite experiments STEP, GG, and MICROSCOPE plan to test the E\"otv\"os parameter $\eta$ to very high accuracy.  This provides a bound on violations of the equivalence principle by quantifying the variation in free-fall acceleration between bodies of differing composition \cite{Will}.  If the chameleon coupling factor $\beta$ is different for different compositions of matter, these differences will manifest themselves in equivalence principle violations which may be detectable by these experiments.
\section{Coda}
\subsection{Summary}
We have given a pedagogical overview of the chameleon field.  We derived its equation of motion from an action and discussed the definition of the matter density and the choice of the potential $V$, showing that a chameleon field with an exponential potential function can account for cosmic acceleration today while remaining hidden from experiments.  Then we discussed the cosmological history of the chameleon field, and finally its interaction with matter today.  With the help of a Maple script, we quickly classified several examples of the chameleon field due to spherical matter distributions and used them to explain not just why the chameleon field has remained hidden from experiments so far but also why it has the potential to be detected.

There are several topics which we did not treat in much detail:  String-theoretical motivations for chameleon fields; the chameleon's transition from the inflation era to the radiation era; equivalence principle violations; and exact constraints on the chameleon model from current experimental data.  Many of these subjects are discussed in \cite{KW} and \cite{BvdBDKW}.

SEE or APSIS will provide strong evidence either for or against the presence of a chameleon field.  If it does exist, many questions will remain:  What is the potential function $V$?  How does the chameleon field behave in the presence of strong gravitational fields?

The chameleon field is an intriguing model because of its promise of explaining cosmic acceleration and because of the possibility that it may soon be tested by satellite experiments.  Even if it is not detected, it may stimulate other ideas for testable alternate theories of gravity, which is an important part of the investigation into the nature of dark energy.
\subsection{Outlook}
The chameleon field has the unusual physical property of density dependence.  Matter density appears in the perfect fluid approximation of other theories, such as Newtonian gravity and bare general relativity, but these theories may be rewritten in terms of enclosed mass or the mass of point particles.  In chameleon gravity, however, there is no obvious way to escape the perfect fluid approximation.  This must be regarded as a shortcoming of the model described here, since in the real Universe the notion of local matter density $\rho$ is not well-defined.  It is essential that the point-particle regime of the chameleon model be explored to see if it is consistent with the perfect fluid approximation.

Even if we accept the perfect fluid approximation, we must remember that the real Universe is inhomogeneous on scales ranging from the sub-atomic all the way up to the cosmological.  Given the non-linear density dependence of the chameleon interaction, we must exercise caution in using the chameleon model to make predictions for real experiments.  A more general solution method or a successful numerical approach to solving the chameleon differential equation would allow us to explore the chameleon on a variety of scales and to make predictions in realistic scenarios rather than just toy models.

In discussing the choice of the potential $V\left(\phi\right)$, we determined that a chameleon field with the power-law potential is ruled out by existing experiments.  But the exponential potential we settled on has the same fine-tuning problem as a bare cosmological constant $\Lambda$, and so it introduces more problems than it solves.  A useful chameleon theory should give $\rho_\phi\sim\rho_\text{m}$ today without resorting to fine-tuning.

The conformal coupling (\ref{conformalrelation}) which we used is not the only possible choice.  Different coupling functions give rise to different chameleon field theories, in which the thin-shell condition may be easier, harder, or impossible, to satisfy \cite{BvdBD}.  In any case, the behaviour of the chameleon force depends strongly on the choice of coupling function, and we have explored only one specific example here.  A different choice may alleviate the problems mentioned above with the potential $V$.

In addition to proposed satellite experiments, chameleon-type modified gravity may have astrophysical effects, for example in the gravitational collapse of gas clouds or in the dynamics of planetary ring systems.  Again, this demands a more general solution method than the approximation we derived in this work.

The recent review by Mota and Shaw \cite{MS}, while not completely answering the above questions, points the way forward, exploring a more general class of potentials than we have done here and also constructing an ``effective macroscopic theory'' of the chameleon field for composite bodies made of small finite-size particles.
\subsection{Acknowledgements}
The author thanks Anne Davis for her supervision and support during the writing of an earlier version of this essay and is grateful to Douglas Scott, Justin Khoury, and Achim Kempf for their constructive feedback and encouragement during the revision process.
\appendix
\makeatletter
\def\@seccntformat#1{Appendix\ \csname the#1\endcsname\quad}
\makeatother
\section{Proof that \texorpdfstring{$m_\text{min}\gg H$}{m\_min>>H}}
\label{proof_mmin_H}
We assume that $\beta$ is of order unity, $n\gtrsim\frac{1}{2}$, and $\phi\left(t\right)$ is slowly-rolling.

Some massaging of equation (\ref{mmin}) will allow us to evaluate $m_\text{min}^2/H^2$.  Start by writing out equation (\ref{phimin}) for this scenario:
\begin{equation}
\begin{gathered}
-nM^n\phi_\text{min}^{-n-1}V\left(\phi_\text{min}\right)+\frac{\beta}{M_\text{pl}}\rho_\text{m}e^{\beta\phi_\text{min}/M_\text{pl}}=0\\
\Rightarrow nM^n\phi_\text{min}^{-n-2}V\left(\phi_\text{min}\right)=\frac{\beta}{\phi_\text{min}M_\text{pl}}\rho_\text{m}e^{\beta\phi_\text{min}/M_\text{pl}}
\end{gathered}
\label{phiminexact}
\end{equation}
Expand the second derivative of the bare potential:
\begin{align*}
V_{,\phi\phi}&=\frac{d}{d\phi}\left(-nM^n\phi^{-n-1}V\right)\\
&=n\left(n+1\right)M^n\phi^{-n-2}V+\left(nM^n\phi^{-n-1}\right)^2V\\
&=\left[\left(n+1\right)+nM^n\phi^{-n}\right]nM^n\phi^{-n-2}V
\end{align*}
Then equation (\ref{mmin}) may be expanded as follows:
\begin{align*}
m_\text{min}^2&=\left[\left(n+1\right)+nM^n\phi_\text{min}^{-n}\right]nM^n\phi_\text{min}^{-n-2}V\left(\phi_\text{min}\right)+\frac{\beta^2}{M_\text{pl}^2}\rho_\text{m}e^{\beta\phi_\text{min}/M_\text{pl}}\\
&=\left[1+n+n\left(\frac{M}{\phi_\text{min}}\right)^n\right]\frac{\beta}{\phi_\text{min}M_\text{pl}}\rho_\text{m}e^{\beta\phi_\text{min}/M_\text{pl}}+\frac{\beta^2}{M_\text{pl}^2}\rho_\text{m}e^{\beta\phi_\text{min}/M_\text{pl}}\\
&=\frac{\beta\rho_\text{m}e^{\beta\phi_\text{min}/M_\text{pl}}}{\phi_\text{min}M_\text{pl}}\left[1+n+n\left(\frac{M}{\phi_\text{min}}\right)^n+\beta\frac{\phi_\text{min}}{M_\text{pl}}\right]
\end{align*}
We can put this together with the Friedmann equation (\ref{Friedmann}) to compute
\begin{equation}
\begin{aligned}
\frac{m_\text{min}^2}{H^2}&=\frac{3M_\text{pl}^2}{\rho_\text{critical}}\frac{\beta\rho_\text{m}e^{\beta\phi_\text{min}/M_\text{pl}}}{\phi_\text{min}M_\text{pl}}\left[1+n+n\left(\frac{M}{\phi_\text{min}}\right)^n+\beta\frac{\phi_\text{min}}{M_\text{pl}}\right]\\
&=3\beta\Omega_\text{m}\frac{M_\text{pl}}{\phi_\text{min}}\left[1+n+n\left(\frac{M}{\phi_\text{min}}\right)^n\right]+3\beta^2\Omega_\text{m}.
\end{aligned}
\label{mminoverHequality}
\end{equation}

Certainly the above equation implies that
\begin{equation}
\frac{m_\text{min}^2}{H^2}>3n\beta\Omega_\text{m}\frac{M_\text{pl}}{\phi_\text{min}}.
\label{mHbound}
\end{equation}
We will consider two cases, $\phi_\text{min}\lesssim M$ and $\phi_\text{min}\gg M$.
 \begin{description}
\item[Case 1:  $\phi_\text{min}\lesssim M$.]We have $\Omega_\text{m}>10^{-28}$ from the Planck era onwards \cite{BvdBDKW}, so equation (\ref{mHbound}) gives
\begin{displaymath}
\frac{m_\text{min}^2}{H^2}\gtrsim3n\beta\Omega_\text{m}\frac{M_\text{pl}}{M}\sim3n\beta\Omega_\text{m}\cdot10^{30}>3n\beta\cdot10^{2}\gg1.
\end{displaymath}
\item[Case 2:  $\phi_\text{min}\gg M$.]Now we have $V\left(\phi_\text{min}\right)\approx M^4$, so equation (\ref{phiminexact}) gives
\begin{gather*}
nM^{4+n}\phi_\text{min}^{-n-1}\approx\frac{\beta}{M_\text{pl}}\rho_\text{m}e^{\beta\phi_\text{min}/M_\text{pl}}=\frac{\beta}{M_\text{pl}}\rho\Omega_\text{m}\\
\Rightarrow\phi_\text{min}^{-1}\approx\left[\frac{\beta}{nM^{4+n}M_\text{pl}}\rho\Omega_\text{m}\right]^\frac{1}{n+1}.
\end{gather*}
Then equation (\ref{mHbound}) becomes
\begin{align*}
\frac{m_\text{min}^2}{H^2}&\gtrsim3n\beta\Omega_\text{m}M_\text{pl}\left[\frac{\beta}{nM^{4+n}M_\text{pl}}\rho\Omega_\text{m}\right]^\frac{1}{n+1}\\
&=3n\beta\left[\frac{\beta}{n}\left(\frac{M_\text{pl}}{M}\right)^n\frac{\rho}{M^4}\Omega_\text{m}^{n+2}\right]^\frac{1}{n+1}.
\end{align*}
From the Friedmann equation (\ref{Friedmann}), we see that as long as $\phi$ is slowly-rolling, $\rho_\text{critical}$ is a decreasing function of time.  Today we have $\rho_\text{critical}\sim M^4$, so from the earliest times until today, we must have $\rho_\text{critical}\gtrsim M^4$, and so
\begin{displaymath}
\frac{m_\text{min}^2}{H^2}\gtrsim3n\beta\left[\frac{\beta}{n}\left(\frac{M_\text{pl}}{M}\right)^n\Omega_\text{m}^{n+2}\right]^\frac{1}{n+1}.
\end{displaymath}
Finally, we need to put a bound on $\Omega_\text{m}$.  In the radiation era and the matter era, $\Omega_\text{m}$ is an increasing function of time, so we'll try to approximate it at the earliest time, $t_M$, found in the $\phi_\text{min}\gg M$ case.  $\phi_\text{min}$ also increases with time as the matter density of the Universe is diluted, so let $t_M$ be the time at which $\phi_\text{min}=M$.  Equation (\ref{phiminexact}) now gives the value of $\rho_\text{m}$ at $t_M$:
\begin{gather*}
nM^nM^{-n-1}V\left(\phi_\text{min}\right)=\frac{\beta}{M_\text{pl}}\rho_\text{m}e^{\beta\phi_\text{min}/M_\text{pl}}\\
\Rightarrow neM^3\approx\frac{\beta}{M_\text{pl}}\rho_\text{m}\\
\Rightarrow \rho_\text{m}\approx\frac{ne}{\beta}M^3M_\text{pl}\sim10^{-16}\,\text{GeV}^4
\end{gather*}
Recall that today we have $\rho_\text{m}\approx\tilde{\rho}_\text{m}\approx10^{-47}\,\text{GeV}^4$.  Assuming that baryonic (and dark) matter had decoupled from the heat bath at $t_M$, the relation $\rho_\text{m}\propto a^{-3}$ then tells us that the ratio of $a$ today to $a$ at $t_M$ is roughly $10^{10}$.  Using $T\propto a^{-1}$ \cite{Peacock} with $T\approx2.73\,\text{K}$ today, we see $T\sim4\,\text{MeV}$ at $t_M$, validating the assumption of decoupling.  All that remains is to compute $\rho_\text{r}=\frac{\pi}{30}gT^4\sim10^{-10}\,\text{GeV}^4$ to see that this era was dominated by radiation, so that $\Omega_{m}\approx\frac{\rho_\text{m}}{\rho_\text{r}}\sim10^{-6}$ at $t_M$.

We conclude that
\begin{align*}
\frac{m_\text{min}^2}{H^2}&\gtrsim3n\beta\left[\frac{\beta}{n}\left(\frac{M_\text{pl}}{M}\right)^n\Omega_\text{m}^{n+2}\right]^\frac{1}{n+1}\\
&\gg3n\beta\left[\frac{\beta}{n}10^{30n}\cdot10^{-6\left(n+2\right)}\right]^\frac{1}{n+1}\\
&\sim10^{\left(24n-12\right)/\left(n+1\right)},
\end{align*}
which is greater than unity for all $n\gtrsim\frac{1}{2}$.
\end{description}
\section{Proof that \texorpdfstring{$\phi_\text{min}\left(t\right)$}{phi\_min(t)} is slowly-rolling}
\label{proof_slowroll}
We assume that $m_\text{min}\gg H$.

We must show that $\dot{\phi}_\text{min}^2/2\ll V$.  Take the time derivative of equation (\ref{phimin}):
\begin{gather*}
V_{,\phi}\left(\phi_\text{min}\right)+\frac{\beta}{M_\text{pl}}\rho_\text{m}e^{\beta\phi_\text{min}/M_\text{pl}}=0\\
\Rightarrow V_{,\phi\phi}\left(\phi_\text{min}\right)\dot{\phi}_\text{min}+\frac{\beta}{M_\text{pl}}\dot{\rho}_\text{m}e^{\beta\phi_\text{min}/M_\text{pl}}+\frac{\beta^2}{M_\text{pl}^2}\rho_\text{m}\dot{\phi}_\text{min}e^{\beta\phi_\text{min}/M_\text{pl}}=0\\
\Rightarrow V_{\text{eff},\phi\phi}\left(\phi_\text{min}\right)\dot{\phi}_\text{min}=-\frac{\dot{\rho}_\text{m}}{\rho_\text{m}}\frac{\beta}{M_\text{pl}}\rho_\text{m}e^{\beta\phi_\text{min}/M_\text{pl}}=-3HV_{,\phi}\left(\phi_\text{min}\right)\\
\Rightarrow\dot{\phi}_\text{min}=-\frac{3HV_{,\phi}\left(\phi_\text{min}\right)}{V_{\text{eff},\phi\phi}\left(\phi_\text{min}\right)}
\end{gather*}
Then
\begin{align*}
\frac{\dot{\phi}_\text{min}^2}{2V\left(\phi_\text{min}\right)}&=\frac{9H^2V_{,\phi}^2\left(\phi_\text{min}\right)}{2V\left(\phi_\text{min}\right)V_{\text{eff},\phi\phi}^2\left(\phi_\text{min}\right)}\\
&<\frac{9}{2}\frac{H^2}{V_{\text{eff},\phi\phi}\left(\phi_\text{min}\right)}\frac{V_{,\phi}^2\left(\phi_\text{min}\right)}{V\left(\phi_\text{min}\right)V_{,\phi\phi}\left(\phi_\text{min}\right)}\\
&=\frac{9}{2}\frac{H^2}{m_\text{min}^2}\frac{V_{,\phi}^2\left(\phi_\text{min}\right)}{V\left(\phi_\text{min}\right)V_{,\phi\phi}\left(\phi_\text{min}\right)},
\end{align*}
where the inequality holds because $V_{\text{eff},\phi\phi}>V_{,\phi\phi}$.  We can also compute
\begin{align*}
\frac{V_{,\phi}^2}{VV_{,\phi\phi}}&=\frac{\left(-nM^n\phi^{-n-1}V\right)^2}{V\left(1+n+nM^n\phi^{-n}\right)nM^n\phi^{-n-2}V}\\
&=\frac{n}{1+n+nM^n\phi^{-n}}\frac{M^n}{\phi^n}\\
&<\frac{n}{nM^n\phi^{-n}}\frac{M^n}{\phi^n}=1.
\end{align*}
And hence if $H^2\ll m_\text{min}^2$, we have
\begin{displaymath}
\frac{\dot{\phi}_\text{min}^2}{2}\ll V.
\end{displaymath}

\begin{lstlisting}[breaklines,upquote=true,label=sphereprofile,caption={[Maple script to solve the static chameleon DE] A Maple script to solve equation (\ref{DE})}]
Digits := 30:
with(linalg):
c := 299792458; # in m/s
Mpl := evalf(1.2e19 / sqrt(8*Pi)); # in GeV
M := 2e-12; # in GeV
V := M^4*exp(M^n/phi^n);
Vphi := diff(V,phi);
Vphiphi := diff(V,phi,phi);
gcm3 := 4.31013337073e-18;
metre := 5.06772886971e15;
printf("Enter a name for this profile:  ");
profilename := readline(terminal);
printf("Enter the value for the coupling constant beta:  ");
beta := parse(readline(terminal));
printf("Enter the value for the exponent n:  ");
n := parse(readline(terminal));
printf("Enter the radius R of the sphere in metres:  ");
R_metre := parse(readline(terminal));
R := R_metre * metre;
printf("Enter the density of the sphere in g/cm^3:  ");
rho_c_gcm3 := parse(readline(terminal));
rho_c := gcm3 * rho_c_gcm3;
printf("Enter the density outside the sphere in g/cm^3:  ");
rho_inf_gcm3 := parse(readline(terminal));
rho_inf := gcm3 * rho_inf_gcm3;
phi_c := fsolve(Vphi + (beta/Mpl)*rho_c = 0, phi, 0..Mpl);
phi_inf := fsolve(Vphi + (beta/Mpl)*rho_inf = 0, phi, 0..Mpl);
m_c := eval(sqrt(Vphiphi + (beta^2/Mpl^2)*rho_c), phi = phi_c);
m_inf := eval(sqrt(Vphiphi + (beta^2/Mpl^2)*rho_inf), phi = phi_inf);

lowcontrastfraction := (phi_inf-phi_c)/(m_c+m_inf+m_c*exp(-2*m_c*R)-m_inf*exp(-2*m_c*R)):
if m_c^2*(lowcontrastfraction*(1+m_inf*R)*(1-exp(-2*m_c*R))/R)<beta*rho_c/Mpl then
printf("Low-contrast condition detected.\n");
A := lowcontrastfraction*(1-m_c*R-exp(-2*m_c*R)-m_c*R*exp(-2*m_c*R));
F := lowcontrastfraction*(1+m_inf*R);
phi := piecewise(r<R,F*(exp(m_c*(r-R))-exp(-m_c*(r+R)))/r+phi_c,r>R,A*exp(-m_inf*(r-R))/r+phi_inf);
closeuprange := R_metre/20;
elif m_c^2*((phi_inf-(1/(1+m_inf*R)+1/2)*beta*rho_c*R^2/(3*Mpl))-phi_c)>beta*rho_c/Mpl then
printf("Thick-shell condition detected.\n");
A := -beta*rho_c*R^3/(3*Mpl*(1+m_inf*R));
phi := piecewise(r<R,beta*rho_c*r^2/(6*Mpl)+(phi_inf-(1/(1+m_inf*R)+1/2)*beta*rho_c*R^2/(3*Mpl)),r>R,A*exp(-m_inf*(r-R))/r+phi_inf);
closeuprange := R_metre/20;
else
printf("Thin-shell condition detected.\n");
continuitymatrix := array( [[0,-1,-X,1-Z],[0,1,0,Y-1+Y*Z+Z],[-1,1,S,0],[T,-1,0,0]] );
solutionmatrix := inverse(continuitymatrix);
X := R_0*phi_c;
Y := m_c*R_0;
Z := exp(-2*Y);
S := R*phi_c;
T := 1+m_inf*R;
continuityrhs := array( [beta*rho_c*R_0^3/(6*Mpl)-phi_c*R_0, beta*rho_c*R_0^3/(3*Mpl), R*phi_inf-beta*rho_c*R^3/(6*Mpl), -beta*rho_c*R^3/(3*Mpl)] );
ACDF := multiply(solutionmatrix,continuityrhs);
R_0_estimate := fsolve(ACDF[4] = (beta*rho_c*R_0)/(m_c^2*Mpl*(1-exp(-2*m_c*R_0))), R_0, 0..R);
R_0 := R_0_estimate;
printf("R_0 = %e\n", R_0);
A := ACDF[1];
phi := piecewise(r<R_0,ACDF[4]*(exp(m_c*(r-R_0))-exp(-m_c*(r+R_0)))/r+phi_c,R_0<r and r<R,beta*rho_c*r^2/(6*Mpl)+ACDF[2]/r+ACDF[3]*phi_c,r>R,ACDF[1]*exp(-m_inf*(r-R))/r+phi_inf);
closeuprange := 2*(R_metre-R_0/metre);
end if:
printf("A = %e\n", A);
W := -A*3*Mpl/(beta*R^3*rho_c);
r := r_metre * metre:
plotsetup(ps,plotoutput=cat(profilename,`phi1.eps`));
plot(phi,r_metre=0..2*R_metre,title=`Phi (GeV)`);
plotsetup(ps,plotoutput=cat(profilename,`phi2.eps`));
plot(phi,r_metre=R_metre-closeuprange..R_metre+closeuprange,title=`Phi (GeV)`);
plotsetup(ps,plotoutput=cat(profilename,`accel1.eps`));
plot((beta/Mpl)*diff(phi,r_metre)*c^2,r_metre=0..2*R_metre,title=`Acceleration (m/s^2)`);
plotsetup(ps,plotoutput=cat(profilename,`accel2.eps`));
plot((beta/Mpl)*diff(phi,r_metre)*c^2,r_metre=R_metre-closeuprange..R_metre+closeuprange,title=`Acceleration (m/s^2)`);

\end{lstlisting}

\begin{thebibliography}{10}
\addcontentsline{toc}{section}{References}
\bibitem{Carroll} S. M. Carroll, ``The Cosmological Constant'', \textit{Living Reviews in Relativity} \textbf{4} (2001):  cited on 23 October 2006, \href{http://www.livingreviews.org/lrr-2001-1}{\mbox{http://www.livingreviews.org/lrr-2001-1}}.
\bibitem{0210533} P. Brax and J. Martin, \href{http://arxiv.org/abs/astro-ph/0210533}{\mbox{astro-ph/0210533}}.
\bibitem{Weinberg} S. Weinberg, Reviews of Modern Physics \textbf{61}, 1--23 (1989).
\bibitem{0403324} V. Sahni, \href{http://arxiv.org/abs/astro-ph/0403324}{\mbox{astro-ph/0403324}}.
\bibitem{KW} J. Khoury and A. Weltman, \href{http://arxiv.org/abs/astro-ph/0309300}{\mbox{astro-ph/0309300}} and \href{http://arxiv.org/abs/astro-ph/0309411}{\mbox{astro-ph/0309411}}.
\bibitem{BvdBD} P. Brax, C. van de Bruck, A.-C. Davis, \href{http://arxiv.org/abs/astro-ph/0408464}{\mbox{astro-ph/0408464}}.
\bibitem{GK} S. S. Gubser and J. Khoury, \href{http://arxiv.org/abs/hep-ph/0405231}{\mbox{hep-ph/0405231}}.
\bibitem{UGK} A. Upadhye, S. S. Gubser and J. Khoury, \href{http://arxiv.org/abs/hep-ph/0608186}{\mbox{hep-ph/0608186}}.
\bibitem{MS} D. F. Mota and D. J. Shaw, \href{http://arxiv.org/abs/hep-ph/0608078}{\mbox{hep-ph/0608078}}.
\bibitem{FM} Y. Fujii and K. Maeda, \textit{The Scalar-Tensor Theory of Gravitation}, University Press, Cambridge (2003).
\bibitem{MTW} C. W. Misner, K. S. Thorne, J. A. Wheeler, \textit{Gravitation}, Freeman, San Francisco (1973).
\bibitem{Damour} T. Damour, \href{http://arxiv.org/abs/gr-qc/9606079}{\mbox{gr-qc/9606079}}.
\bibitem{Ratra_Peebles} B. Ratra and P. J. E. Peebles, Physical Review D \textbf{37}, 3406--27 (1988).
\bibitem{PB} P. Bin\'etruy, International Journal of Theoretical Physics \textbf{39}, 1859--75 (2000).
\bibitem{ZWS} I. Zlatev, L. Wang, P. J. Steinhardt, Physical Review Letters \textbf{82}, 896--9 (1999).
\bibitem{BvdBDKW} P. Brax, C. van de Bruck, A.-C. Davis, J. Khoury, A. Weltman, \href{http://arxiv.org/abs/astro-ph/0408415}{\mbox{astro-ph/0408415}} and \href{http://arxiv.org/abs/astro-ph/0410103}{\mbox{astro-ph/0410103}}.
\bibitem{WMAP3} D. N. Spergel et al., \href{http://arxiv.org/abs/astro-ph/0603449}{\mbox{astro-ph/0603449}}.
\bibitem{plog} E. W. Weisstein, ``Lambert W-Function'',  \href{http://mathworld.wolfram.com/LambertW-Function.html}{\mbox{http://mathworld.wolfram.com/LambertW-Function.html}}.
\bibitem{AHN} E. G. Adelberger, B. R. Heckel, A. E. Nelson, Annual Review of Nuclear and Particle Science \textbf{53}, 77--121 (2003).
\bibitem{Peacock} J. A. Peacock, \textit{Cosmological Physics}, University Press, Cambridge (2005).
\bibitem{Eddington} A. S. Eddington, Proceedings of the Royal Society of London A \textbf{111}, 424--56 (1926).
\bibitem{HNSS} J. K. Hoskins, R. D. Newman, R. Spero, J. Schultz, Physical Review D \textbf{32}, 3084--95 (1985).
\bibitem{SEE1} A. J. Sanders et al., Measurement Science and Technology \textbf{10}, 514--524 (1999).
\bibitem{SEE2} A. D. Alexeev et al., \href{http://arxiv.org/abs/gr-qc/0002088}{\mbox{gr-qc/0002088}}.
\bibitem{APSIS} V. Sahni and Y. Shtanov, \href{http://arxiv.org/abs/gr-qc/0606063}{\mbox{gr-qc/0606063}}.
\bibitem{Will} C. M. Will, \textit{Theory and experiment in gravitational physics}, University Press, Cambridge (1981).
\end{thebibliography}
\end{document}